\documentclass[11pt]{article}
\usepackage[utf8]{inputenc}
\usepackage[margin=1in]{geometry}
\usepackage{amsmath}
\usepackage{amssymb}
\usepackage{amsthm}
\usepackage{graphicx}
\usepackage{setspace}
\usepackage{enumitem}
\usepackage{float}
\usepackage{xcolor}
\usepackage{natbib}
\usepackage{booktabs}
\usepackage{tabularx}
\usepackage{array}
\usepackage{authblk}
\usepackage[hidelinks]{hyperref}

\title{Conspiracy Theory Rabbit Holes Emerge via Interacting Contagions}
\author[1]{Fabian Tschofenig}
\author[2]{Paul Vicinanza}
\author[3]{Henrich Greve}
\author[1]{Hayagreeva Rao}
\author[1]{Douglas Guilbeault\thanks{Corresponding author: \href{mailto:dguilb@stanford.edu}{dguilb@stanford.edu}}}

\affil[1]{Stanford Graduate School of Business}
\affil[2]{Amazon}
\affil[3]{INSEAD Singapore}
\date{}

\begin{document}
\maketitle

\begin{abstract}
\noindent Why do people fall into conspiracy theory rabbit holes? Prior research explains rabbit holes via psychological and algorithmic causes, yielding inconsistent findings. Here, we argue that rabbit holes can also arise from interactions among conspiracy theories spreading as social contagions. Using 7.6 million tweets from 7,416 X users during the first wave of COVID-19, we identify public endorsement of 15 conspiracy narratives with prompt-tuned large language models. Sequential hazard models show that, characteristic of rabbit holes, adopting a conspiracy theory elevates the risk of sharing subsequent conspiracy theories, that this elevation grows and persists longer with the number of conspiracy theories shared, and that transitions between theories concentrate among semantically proximate narratives, revealing semantic interactions that mediate social contagions. We also document what we term the settler effect: a user's entry into a new semantic region is slower, but once entry occurs, subsequent within-region adoption accelerates. We compare a range of agent-based models in their ability to reproduce these dynamics. Neither independent adoption nor a generic post-adoption increase in susceptibility reproduces the joint temporal and semantic pattern of the settler effect; among the alternatives considered, an ecology-of-contagions model that formalizes belief-system reshaping most parsimoniously reproduces these patterns. Using counterfactual network simulations that account for interactions among conspiracy theories, we find that preventing the first public endorsement of a conspiracy theory can rival high-detection shadow banning and outperform week-long read-only lockouts at reducing the spread of conspiracy theories.
\end{abstract}
\newpage
\section{Introduction}

The impact of conspiracy theories can be devastating. During the COVID-19 pandemic, belief in such claims reduced preventive behaviours and vaccination intentions \citep{romer2020, loomba2021}, ultimately costing lives \citep{jia2023estimated, bollenbacher2026effects}. Beyond the pandemic, descent into conspiracy thinking has been linked to vaccine refusal, racial prejudice, and political violence \citep{ jolley2014effects, jolley2020exposure, jolley2020pylons}, and once these beliefs are entrenched, they prove resistant to correction \citep{cichocka2020counter, carey2022ephemeral}. How that endorsement spreads, from person to person and from theory to theory, is both a scientific puzzle and a public concern.

Conspiracy theories, explanations that attribute events to the secret coordination of powerful actors \citep{difonzo2018conspiracy}, are seldom endorsed in isolation \citep{greve2022}. People who endorse one tend to endorse others, including mutually contradictory ones \citep{goertzel1994, wood2012}, a tendency towards clustering \citet{goertzel1994} described as a monological belief system characterized by a generalized skepticism. Stable individual predispositions partly account for who engages with conspiracy theories in the first place \citep{imhoff2014, brotherton2013}, and a separate line of work has identified gateway theories — relatively plausible narratives such as distrust in mainstream media or skepticism about official COVID-19 case counts — that disproportionately appear as users' first endorsement before progression to more extreme claims \citep{greve2022}. These accounts tell us who is at risk and which narratives tend to come first. They tell us much less about whether prior endorsement actually raises susceptibility to the next one, or whether the pattern is mere co-occurrence among the predisposed. If there is an effect, its structure is also unresolved: namely, whether some theories elevate susceptibility more than others, whether the elevation is lasting or transient, and how peer exposure shapes the process when multiple people in a user's network are sharing conspiracy content.

The sequential pattern of conspiracy theory adoption, often described as a “rabbit hole,” refers to the observed trajectory in which users adopt one conspiracy theory and then continue adopting additional theories with increasing speed and breadth. This metaphor has become dominant in both popular and academic discussion of online radicalization, yet \citet{sutton2022rabbit} found that almost no research has examined whether or why users actually descend these holes. Early accounts attributed the rabbit hole trajectory to algorithmic radicalization, but that view has weakened: users descend despite platform guardrails \citep{chen2023subscriptions, liu2025short}, large-scale exposure interventions on Facebook do not reduce extremist belief \citep{guess2023social, nyhan2023likeminded, allcott2024effects}, and extreme content is more often sought out by individuals rather than algorithmically supplied \citep{robertson2023users}. However, a shortcoming of the rabbit hole metaphor -- stemming from its source in \textit{Alice in Wonderland} -- is its individualist frame, describing the descent of a single person encountering content in isolation. But given the extent to which content exposures online are driven by social interactions, rabbit holes can be productively viewed as a collective phenomenon that emerges through the interaction of people and conspiracy theories. We therefore approach these rabbit hole trajectories through the lens of interacting social contagions.

Users do not generate conspiracy theories in isolation; they encounter them because someone else in their network has shared, retweeted, or replied to them, and they in turn pass them on to others. This is the defining structure of social contagion. Yet research on conspiracy belief has only sparingly engaged the contagion literature, emphasizing instead psychological needs, network exposure as a static feature, and stable predispositions. Given the credibility costs and reputational risk that publicly endorsing a conspiracy theory carries, their spread is expected to show the patterns and properties typically studied as complex contagions, where adoption requires reinforcement from multiple independent sources rather than a single exposure \citep{centola2007complex, guilbeault2018complex}. But complex contagion theory typically describes one contagion at a time. In a recent review, \citet{hebert-dufresne2025} call for more formal and empirical research into what Guilbeault et al. (2018) refer to as the ``ecology of contagions,'' namely how systematic interactions among contagions shape their diffusion patterns. Addressing this gap is especially important in the context of studying conspiracy theory adoption, since rabbit hole descent suggests that what matters most about conspiracy theories is how they interact with one another. To our knowledge, no empirical study has treated conspiracy theories as interacting contagions in both a theoretically clear and empirically validated manner. This is due in part to the methodological challenges in measuring how social contagions interact, not only behaviorally via exposures, but also semantically in the context of narratives and related online content. 


We argue that rabbit holes are not simply sequences of independent adoptions; they are the observable outcome of interacting contagions. This framing shifts the empirical question from whether users adopt multiple conspiracy theories to how the adoption of one narrative changes susceptibility to others via interactions in social networks: which narratives become easier to adopt, whether these effects persist or decay, and how peer exposure amplifies them. Existing models, however, rarely scale beyond two contagions. As the number of contagions increases, the number of possible interactions grows rapidly, making the parameter space difficult to identify empirically \citep{hebert-dufresne2020, hebert-dufresne2025}. We address this problem by treating conspiracy theories not as isolated contagions, but as part of an ecology of contagions \citep{guilbeault2018complex}: a system of co-circulating narratives whose interactions are mediated by their semantic relatedness and by the belief systems of the individuals who encounter and share them.

In this ecology metaphor, contagions need not interact directly per se, since they primarily interact through users. Each narrative encodes assumptions about trust, evidence, agency, threat, and deception, and each user evaluates new narratives through their existing interpretative stance, which can be described as a function of their beliefs and semantic associations \citep{boutyline2017belief}. For simplicity and consistency, we will adopt the language of ``beliefs'' throughout, since this is common to the empirical study of conspiracy theories, though we note that belief dynamics can also be understood through the lens of associative reasoning via interpretations and semantic priming \citep{olsson2024analogies}. In our ecology-of-contagion view, adoption is driven by a compatibility judgment between the user’s current belief state and the narrative content encoded by the narrative. Indeed, prior work (e.g., on cognitive dissonance, confirmation bias, and prototype fluency) shows that people are more likely to accept narratives that fit their existing category frameworks and belief systems and less likely to embrace those that do not \citep{festinger1962cognitive, zuckerman1999categorical, winkielman2006prototypes, mercier2017enigma}. We theorize that narrative adoption shifts a user's belief system toward the adopted narrative, thereby changing the compatibility of subsequently encountered narratives. Interactions among contagions can thus arise indirectly, through the way each adoption alters the evaluative context for the next, yielding a kind of associative diffusion across contagions \citep{goldberg2018beyond, aiyappa2024emergence}.

This framework requires an empirical proxy for the latent belief system through which users evaluate these narratives, since it is intrinsically challenging to directly measure people's underlying interpretative stance and belief system from their public displays of communication and information consumption. Prior work on belief systems motivates treating beliefs as interdependently structured rather than independent \citep{boutyline2017belief}, while recent work shows that semantic embeddings can approximate latent ideological or belief spaces from language \citep{lee2025}. This motivates our use of semantic embeddings as a proxy for this latent belief space.

Two further problems apply directly to studying conspiracy theories. A core methodological challenge is that conspiracy theories circulate as narratives, not as fixed strings of text. The same conspiracy theory can be expressed through different hashtags, slogans, insinuations, and rhetorical formulations, while many of the same words can be used to endorse, reject, mock, or simply discuss a claim. This creates a key methodological limitation for keyword- and hashtag-based approaches: they can track visible markers of conspiracy talk, but they cannot reliably recover the narrative being endorsed or distinguish endorsement from mere exposure or discussion.
Rather than imposing a fixed dictionary of terms, we first identify the recurring conspiracy-theory narratives that appear in the corpus. We then prompt-tune large language models to identify tweet-level endorsements of each narrative, while avoiding false positives from tweets that merely mention, criticize, or discuss the narrative. The classifiers distinguish endorsement from mention, criticism, and adjacent discussion, and they reproduce human labels with high accuracy. This makes it possible to measure genuine endorsement of conspiracy theories across millions of tweets, instead of relying only on the presence of particular words, hashtags, or URLs (Materials and Methods; Section \ref{apx:sec_data_processing_pipeline} of the Supplementary Appendix).


A second challenge is to distinguish interaction among contagions from how their shared adoption is shaped by stable, underlying predispositions. In observational data, contagions that co-occur at the population level may be genuinely interacting, or they may reflect shared individual-level predispositions rather than interaction between the contagions themselves, a general problem in inferring interaction from co-occurrence \citep{blanchet2020cooccurrence}. We address this challenge by comparing three minimal agent-based models under a shared illustrative synthetic initialization: independent adoption, a generic post-adoption increase in susceptibility, and semantically structured reshaping, in which prior adoption raises susceptibility as a function of semantic proximity. This third model provides a minimal realization of an ecology of contagions. As we show in the analyses to follow, only this model reproduces the full joint empirical signatures of conspiracy theory adoption we report, moving the evidence beyond co-occurrence and toward genuine interaction. We also distinguish the process that brings users into a conspiracy theory from the process that sustains their subsequent sharing. We model first adoption as a hazard process and subsequent re-sharing as a self-exciting Hawkes process \citep{hawkes1971spectra, zarezade2017correlated}. This separates the question of when users first endorse a conspiracy theory from the bursty dynamics through which they continue to share it, which homogeneous and inhomogeneous Poisson models fail to reproduce.

We test this argument using 7.6 million tweets from 7,416 X users during the first wave of the COVID-19 pandemic \citep{greve2022}. Tweets are classified with prompt-tuned language models that identify endorsement rather than mere mention, criticism, or adjacent misinformation. We reconstruct each user’s sequence of first observed endorsements across 15 identified conspiracy theories and estimate sequential Cox proportional hazard models with peer exposure as a time-varying covariate. These models track how the hazard of subsequent adoption changes with prior adoption, semantic proximity, and recent sharing by neighbours. We then compare the observed temporal and semantic structure of adoption with minimal alternative models and use the fitted dynamics in an agent-based simulation on the observed network to examine intervention counterfactuals.

\section{Empirical findings}\label{sec:empirical}
To test whether conspiracy theories spread as interacting rather than independent contagions, we follow users' public endorsements of multiple conspiracy-theory narratives over time. We treat each conspiracy theory as a contagion and define adoption as the first observed tweet in which a user publicly endorses that narrative. This allows us to reconstruct how users move from one conspiracy theory to another across the observation period.

The analysis uses 7.6 million tweets posted by 7,416 users during the first months of the COVID-19 pandemic and related crisis events in 2020. Users enter the sample through two prominent COVID-19 conspiracy hashtags, \textit{\#coronahoax} and \textit{\#virushoax}, but adoption events are measured from their broader tweet histories rather than only from tweets containing those seed hashtags.

Rather than assuming a fixed set of conspiracy theories in advance, we first identify the recurring narratives that appear in the corpus. We then measure whether users endorse each narrative, distinguishing endorsement from mere mention, criticism, or discussion. Endorsement classifications are validated against human labeling to ensure that the measures capture genuine endorsement rather than simple reference to a narrative. For full details, see the Materials and Methods section and section \ref{apx:sec_data_processing_pipeline} of the supplementary appendix. 

This procedure yields the conspiracy-theory narratives reported in Table \ref{tab:categories}. The table lists 16 identified narratives, but the analyses use 15 adoption categories because the Anthony Fauci and Hydroxychloroquine narratives are combined. These narratives are difficult to separate empirically: Fauci is central to many Hydroxychloroquine claims, and the two occupy a highly overlapping region of semantic space.

\begin{table}[H]
\centering
\caption{Identified conspiracy theory narratives.}
\label{tab:categories}

\begingroup
\small
\setstretch{1}
\setlength{\tabcolsep}{4pt}
\renewcommand{\arraystretch}{0.9}

\begin{tabularx}{\textwidth}{@{}>{\raggedright\arraybackslash}p{0.22\textwidth}>{\raggedright\arraybackslash}X@{}}
\toprule
\textbf{Conspiracy theory} & \textbf{Description} \\
\midrule
5G & 5G radio towers are a tool for brain control or cause mass illness and sterilization, are the secret cause of COVID-19. \\
Anti Vaccine & Vaccines contain tracking/mind control microchips or are a tool for forced sterilization/genocide. COVID-19 is a ploy to push vaccines. \\
Bill Gates & Bill Gates is a bioterrorist seeking to depopulate the planet through a variety of means, including vaccines or developing COVID-19. \\
Black Lives Matter & The BLM protests are a tool of antifa/democrats/globalists/George Soros to gain power / usher in a socialist revolution. \\
China Created Covid & The COVID-19 virus was created in a Chinese laboratory and deliberately released as a bioweapon. \\
Democrats & The Democratic elite act as a secret cabal to manipulate votes/destroy the country. Used the pandemic to crash the economy for political gain. \\
Empty Hospitals & The hospitals during COVID-19 had no COVID patients. They were secretly empty or full of paid actors. \\
Fake News & ``Mainstream'' media deliberately spreads false or misleading information to pursue a covert agenda. \\
False Positive Testing & COVID-19 tests are rigged to give false positive results and exaggerate the severity of the pandemic. \\
Anthony Fauci & Anthony Fauci planned the pandemic (e.g., developed the virus), is profiting off the virus, or is using it to pursue nefarious means. \\
Hydroxy-chloroquine & Hydroxychloroquine is a secret cure for COVID-19 that is being suppressed by the CDC/Big-pharma for profit or to inject vaccines. \\
Miscounted Deaths & Hospitals are miscounting deaths from car accidents or heart attacks as COVID-19 deaths for federal funding or to exaggerate the pandemic. \\
Pizzagate & The Democratic party, global elites, and Hollywood run a secret human trafficking and child sex ring. They may sacrifice children and drink their blood. \\
Plandemic & A viral video that proffers the entire pandemic is staged by a secretive cabal. \\
QAnon & Donald Trump is waging a war against the shadowy deep state and powerful pedophiles. \\
Trump Puppet & Vladimir Putin secretly controls Donald Trump and the Republican party, either through blackmail or payment. \end{tabularx}
\endgroup
\end{table}
To analyze the effect of semantic similarity between conspiracy theories on adoption dynamics, we also group the identified narratives into broader semantic clusters. These regions summarize which narratives are closer to or farther from one another in the corpus and allow us to distinguish adoption within a semantic region from movement across regions. For full details on the construction of these regions, see the Materials and Methods section and section \ref{apx:sec_hierarchical_clustering} of the supplementary appendix.

The empirical analysis then follows users' observed adoption sequences. For each user, we record which conspiracy-theory narrative they publicly endorse first, which one they endorse next, and how this sequence unfolds over time. We define adoption depth as the number of distinct conspiracy theories a user has adopted at a given point in the sequence: zero before any observed adoption, one after the first adoption, two after the second, and so forth. We first examine peer exposure, asking whether reinforcement from multiple sharing neighbors operates differently before and after a user enters the adoption sequence. We then analyze within-individual event histories to test whether prior adoption increases the hazard and persistence of further adoption, and whether escalation is content-specific, structured by semantic proximity, and shaped by movement within or across broader semantic regions.

\subsection{The Effects of Social Reinforcement on Conspiracy Theory Adoption}

Conspiracy-theory adoption requires exposure, and in our setting exposure is measured through recent sharing by network neighbours. The ecology-of-contagions framework predicts that the effect of exposure may depend on where a user is in the adoption sequence: before first adoption, exposure determines whether a conspiracy theory reaches the user; after first adoption, exposure may also interact with the user's prior adoption history.

We test this by estimating how the adoption hazard changes with the number of distinct neighbours who shared the target conspiracy theory during the prior 14 days. For first adoption, the effect of additional sharing neighbours rises initially but quickly plateaus. For later adoptions, the hazard increases more steadily as the number of sharing neighbours grows, reaching almost double the one-neighbour baseline when four or more neighbours shared the target theory (Fig.~\ref{fig:hazard_panel}a). Thus, social reinforcement takes on the cumulative shape expected of complex contagion mainly after users have already entered the adoption sequence (i.e., the rabbit hole).

This result shows that between-individual transmission is state-dependent. Peer exposure helps determine which conspiracy theories reach users, but its relationship to adoption changes after prior adoption. We next turn to the within-individual side of the process: whether adopting one conspiracy theory is followed by elevated susceptibility to adopting others.

\subsection{Adoption Elevates and Prolongs Susceptibility.}
If conspiracy theories spread independently, adopting one theory should not systematically alter the subsequent risk of adopting another after accounting for recent peer exposure. If conspiracy theories really behave like interacting contagions, we expect to see a post-adoption trace in the event history: later adoption should become more likely after a prior adoption, and this elevation should potentially grow as users accumulate adopted theories.

We test this by estimating sequential hazard models in which the first-adoption baseline captures entry into the adoption sequence and, for later adoptions, the event clock resets at the user's most recent prior adoption. The models include time-varying reinforcement from sharing neighbours, so the comparison captures post-adoption changes in susceptibility conditional on measured recent peer exposure.

\begin{figure}[H]
    \centering
    \includegraphics[width=\textwidth]{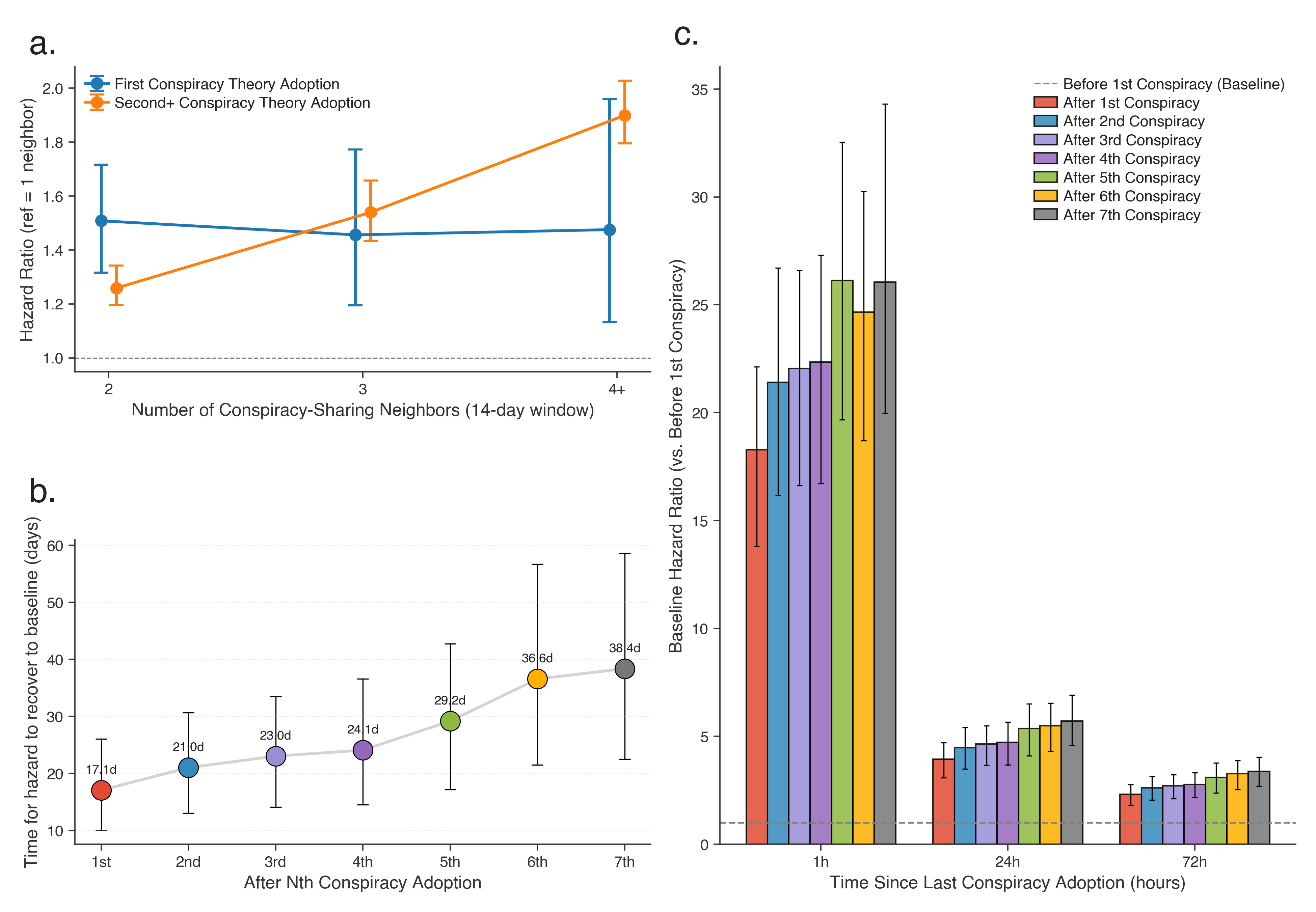}
    \caption{\textbf{Exposure and post-adoption hazard dynamics.} \textit{(a)} Hazard ratios for first adoption and second or later adoption as a function of sharing neighbors within a 14-day window, referenced to one exposed neighbor. Error bars: 95\% CIs. \textit{(b)} Time for the post-adoption hazard to recover to the first-adoption baseline by adoption depth. \textit{(c)} Baseline hazard ratios at 1~h, 24~h, and 72~h after the most recent adoption, relative to the never-adopter baseline. Error bars: 95\% CIs from user clustered bootstrapping}
    \label{fig:hazard_panel}
\end{figure}

The post-adoption hazard is sharply elevated and increases with adoption depth (Fig.~\ref{fig:hazard_panel}c). Immediately after first adoption, the hazard of adopting another conspiracy theory is nearly 20 times the never-adopter baseline; the elevation remains visible at 24 and 72 hours and increases with the number of conspiracy theories adopted. The elevated-risk period also lengthens with depth (Fig.~\ref{fig:hazard_panel}b): the recovery window expands from about 17 days after first adoption to roughly 38 days after the seventh. Full depth-specific estimates are shown in Fig.~\ref{fig:hazard_panel}.

These results show that adoption is followed by elevated later susceptibility, and that both the level and persistence of this elevation increase with adoption depth. Users who move deeper into the sequence are not merely high-propensity adopters experiencing identical bursts of activity after each endorsement. Rather, each adoption is followed by a period of elevated risk, and this period becomes stronger and longer as adoption depth increases.

\subsection{Semantic Proximity Mediates Sequences of Conspiracy Theory Adoption}
The preceding results show that adoption is followed by elevated later susceptibility. We next ask whether this susceptibility is structured by the semantic organization of conspiracy theories. In the ecology-of-contagions framework, conspiracy theories occupy positions in a shared semantic space. If prior adoption reshapes a user's belief system, later adoption should concentrate among narratives that are semantically compatible with what the user has already endorsed rather than rising uniformly across all conspiracy theories. Semantic distance therefore provides an observable proxy for testing this account.

\begin{figure}[H]
    \centering
    \includegraphics[width=\textwidth]{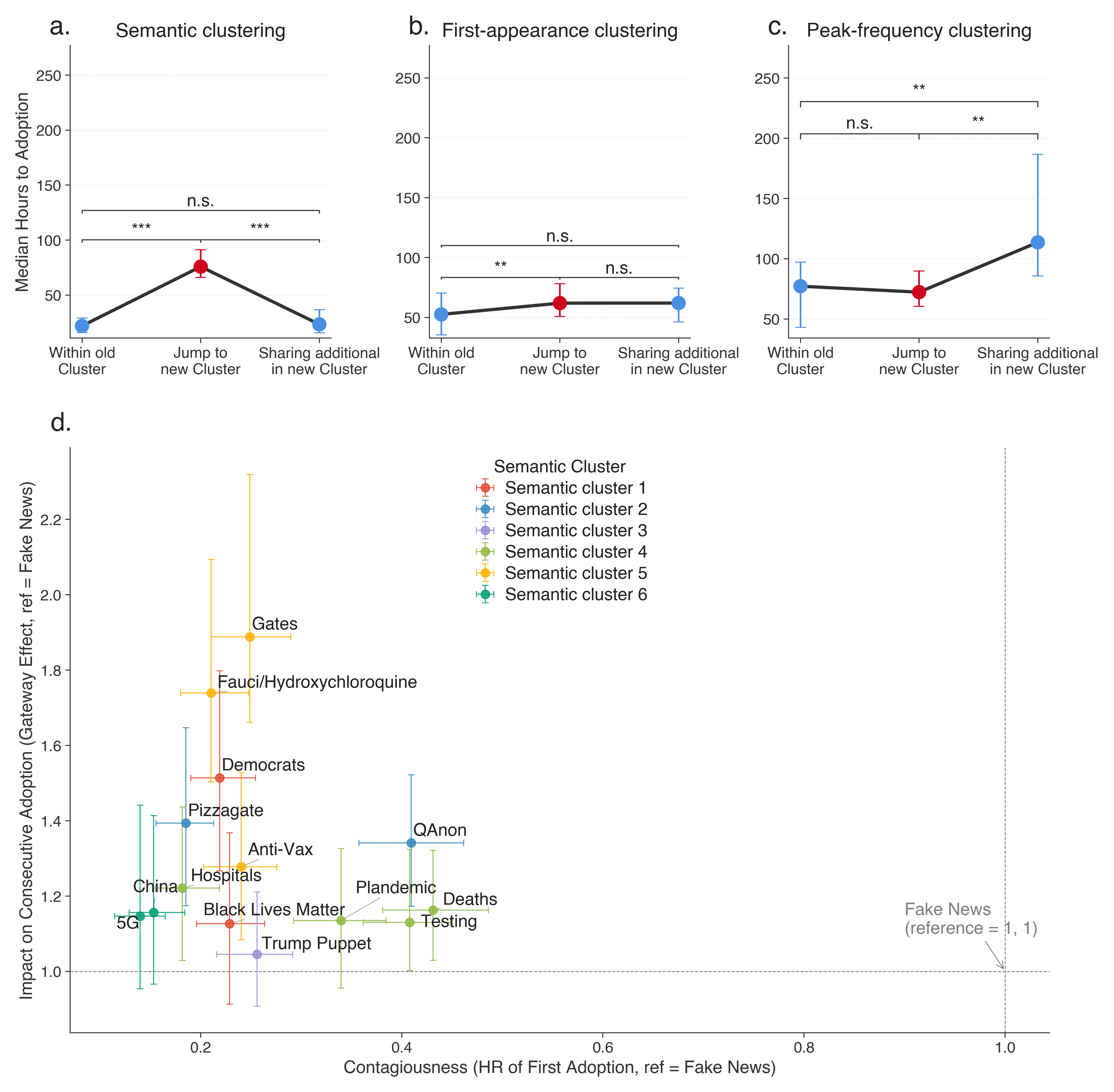}
    \caption{\textbf{Content-specific facilitation and potency.} \textit{(a)} Semantic clustering, \textit{(b)} first-appearance clustering, and \textit{(c)} peak-frequency clustering show median hours to next adoption for within-cluster, jump, and settle transitions. Error bars: 95\% bootstrap CIs; brackets: Holm-corrected pairwise Wilcoxon tests. \textit{(d)} Contagiousness versus potency for each conspiracy theory, relative to \textit{Fake~News}; colour indicates semantic cluster.}
    \label{fig:settler_gateway}
\end{figure}

This is what we observe. Conspiracy theories that are closer in semantic space are also more likely to appear together in users' adoption histories. We compare the semantic-distance matrix with a user-level co-adoption matrix based on pairwise Jaccard distance across users and find a moderate but reliable association between the two (Mantel Spearman correlation = 0.47, $p = 0.0001$). Semantic structure also appears in the timing of adoption. We derive semantic regions by applying hierarchical clustering to the conspiracy-theory distance matrix (Material and Methods). Users move substantially faster among conspiracy theories in semantic regions they have already entered than into regions they have not previously visited (Fig.~\ref{fig:settler_gateway}a). In the sequential hazard models, adoption from an unvisited semantic region occurs at roughly half the hazard of adoption from a previously entered region, after controlling for peer exposure, degree and conspiracy-theory fixed effects.

This structure produces a pattern we term the settler effect. First entry into a new semantic region is slow, but once a user has entered that region, later adoption within it accelerates (Fig.~\ref{fig:settler_gateway}a). The settler effect appears under semantic clustering (Fig.~\ref{fig:settler_gateway}a) but not under temporal null clusterings based on first appearance or peak adoption frequency (Fig.~\ref{fig:settler_gateway}b,c), indicating that semantic proximity captures structure in adoption sequences that simple temporal co-circulation does not. 

Finally, the conspiracy theories that draw users into the adoption sequence are not the same as those that most strongly accelerate subsequent adoption. We define contagiousness as a conspiracy theory’s relative cause specific hazard of being adopted first, and potency as the conditional association between having adopted that theory first and the hazard of adopting another conspiracy theory next, with Fake News as the reference category. Fake News is the most common first adoption but is not especially potent. Several less common first adoptions, including Bill Gates and the Fauci/Hydroxychloroquine theory, are associated with higher second adoption risk (Fig.~\ref{fig:settler_gateway}d). The absence of conspiracy theories that are both highly contagious and highly potent suggests that easy entry and downstream acceleration are distinct roles in the ecology of COVID-19 conspiracy theory contagions observed here.

Together, these results indicate that susceptibility following adoption is not uniform but is instead organized by the geometry of semantic space. Later adoption concentrates among narratives that are semantically proximate to those already endorsed. The settler effect gives this pattern a specific temporal signature: slow entry into a new semantic region followed by accelerated adoption within it. The contagiousness--potency distinction further shows that the theories most likely to draw users into the adoption sequence are not necessarily those associated with the greatest increase in subsequent adoption risk. 

\section{Formal Model Comparisons}

The empirical results show that conspiracy-theory adoption sequences have both a temporal dynamic and a semantic structure. Adoption raises the risk of later adoption, this elevated risk compounds and persists with adoption depth, and later adoption concentrates among semantically related narratives. These are the empirical signatures of rabbit-hole dynamics. We argue, however, that the rabbit hole itself is not the mechanism. It is the observed trajectory produced when prior adoption changes the conditions under which later narratives become adoptable --- in other words a symptom of the interactions between conspiracy theories.

We argue that the mechanism is an ecology of contagions: conspiracy theories interact through the belief systems of the individuals who encounter them. We formalize this account in the Ecology of Contagions Model, where adoption reshapes the user's belief-system state toward the adopted narrative and thereby changes which other narratives become more compatible next. To clarify this mechanism, we compare three minimal models in terms of their ability to reproduce the observed empirical patterns of cumulative conspiracy adoption: the \textit{No Interaction Model}, the \textit{General Arousal Model}, and the \textit{Ecology of Contagions Model}. The formal equations, parameter definitions, simulation rules, and illustrative simulations appear in the Methods and Appendix \ref{sec:formal_model_appendix}. In what follows, we describe the main results of our analyses. 

All three models share the same baseline adoption process. Users' belief systems and conspiracy theories are represented in a shared latent (and low dimensional) belief-system space, with semantic distance serving as a proxy for compatibility. In all three models, conspiracy theories closer to the user's current belief-system state are more likely to be adopted than the ones farther away. The models differ only in what happens after adoption. In the \textit{No Interaction Model}, adoption has no altering effect. In the \textit{General Arousal Model}, adoption produces a temporary, content-neutral increase in susceptibility to all conspiracy narratives, capturing explanations based on heightened attention, increased interest in conspiracy content, or bursts of platform activity. In the \textit{Ecology of Contagions Model}, adoption reshapes the user's belief-system state by shifting it toward the adopted narrative, thereby changing which other narratives become more compatible next.

The \textit{No Interaction Model} can generate co-adoption, but not descent. Some users may adopt many conspiracy theories because they begin close to multiple narratives in belief-system space or because they have high baseline susceptibility. But adoption itself has no effect on what happens next. Because adoption itself has no effect, the model cannot generate a post-adoption rise in hazard, a depth-dependent increase in susceptibility, or a recovery window that expands with repeated adoption. The shared static geometry can still produce faster within-cluster transitions, but it does not generate adoption-dependent semantic reshaping or the settler pattern.

The \textit{General Arousal Model} can generate descent, but not semantic structure. Because adoption temporarily raises susceptibility to all conspiracy theories, users who enter the sequence become more likely to continue adopting. This can produce a rabbit-hole-like acceleration after entry. But the trajectory is undifferentiated: any conspiracy theory becomes more likely, regardless of its relation to what the user has already adopted. The model therefore cannot reproduce the semantic concentration of later adoption, the faster transitions within semantic clusters, the slow jump into a new semantic region followed by faster settling, or the dissociation between contagiousness and potency.

The \textit{Ecology of Contagions Model} produces structured descent. In the model, adoption raises later susceptibility because it reshapes the user's modeled belief system state, but the increase is not uniform. It is strongest for narratives that become more compatible with the reshaped belief-system state. The accumulation of adopted narratives can generate the compounding and persistence documented in our empirical analyses, while semantic distance organizes which narratives become likely next. The same logic generates the settler pattern: first entry into a new semantic region is harder because it requires a larger belief-system adjustment, but once a narrative in that region is adopted, the belief system shifts toward that region and later adoption within it becomes faster.

The \textit{No Interaction Model} can produce co-adoption without descent, while the \textit{General Arousal Model} can produce descent without semantic reshaping caused by adoption. The \textit{Ecology of Contagions Model} produces structured descent by formalizing belief-system reshaping: adoption of one contagion shifts the user’s modeled belief state, which then changes the relative adoption probabilities of other contagions. This does not establish experimental causality or directly observe private belief change. Rather, it shows that the joint temporal and semantic signatures observed in the endorsement histories are consistent with the belief-system reshaping mechanism implied by an ecology of contagions.

These comparisons are not intended as a full psychological model of belief change, nor do they directly identify private belief states. Rather, they clarify whether belief-system reshaping provides a minimal mechanism capable of generating the joint temporal and semantic signatures observed in the endorsement histories, relative to simpler alternatives based on independent adoption or general arousal.

\section{Agent-based Network Simulation and Policy Counterfactuals}\label{sec:simulation}

The empirical results suggest that adoption has two consequences: it creates a public signal that can expose others, and it changes the adopter's susceptibility to other conspiracy theories. We offer a proof-of-concept for how these insights can be leveraged to enhance policy strategy directed as conspiracy theory prevention. Specifically, we use an agent-based simulation to examine hypothetical counterfactual scenarios, based on the current literature, for how policy interventions perform in light of the identified features of conspiracy belief spreading dynamics.

The simulation embeds the fitted sequential hazard models in the observed human-only retweet and reply network. At each simulation step, users face risks of adopting conspiracy theories they have not yet shared; these risks are determined by the fitted hazard models, including adoption depth, peer exposure and semantic clusters. After adoption, repeat sharing follows a fitted Hawkes process of self excitation, allowing adoption events to generate the additional exposure for neighbours. Appendix Figure~\ref{fig:appendix_hawkes} validates this choice by showing that the Hawkes process is strongly preferred by AIC over homogeneous and inhomogeneous Poisson alternatives for empirical repeat sharing sequences. Across the 100 baseline runs, the ABM closely reproduces the empirical prevalence of conspiracy theory adoption and the distribution of first-adoption entry points, while also capturing the observed coadoption structure across conspiracy theories well. Full validation diagnostics are reported in section \ref{apx:sec_simulation_validation} of the Supplementary Appendix. 

We compare three interventions that target different points in the process. Quarantine temporarily prevents posting after adoption, shadow banning reduces the visibility of conspiracy-related posts to neighbours, and reputation nudges intervene before public endorsement. We compare three counterfactual intervention scenarios using the area under the adoption curve (AUC), defined as the integral over time of the share of users who have shared each conspiracy theory. AUC is high when sharing rises quickly and reaches many users, and low when sharing is slow, limited or both. Each intervention is evaluated as the percentage reduction in AUC relative to the baseline simulation.

\begin{figure}[H]
    \centering
    \includegraphics[width=\textwidth]{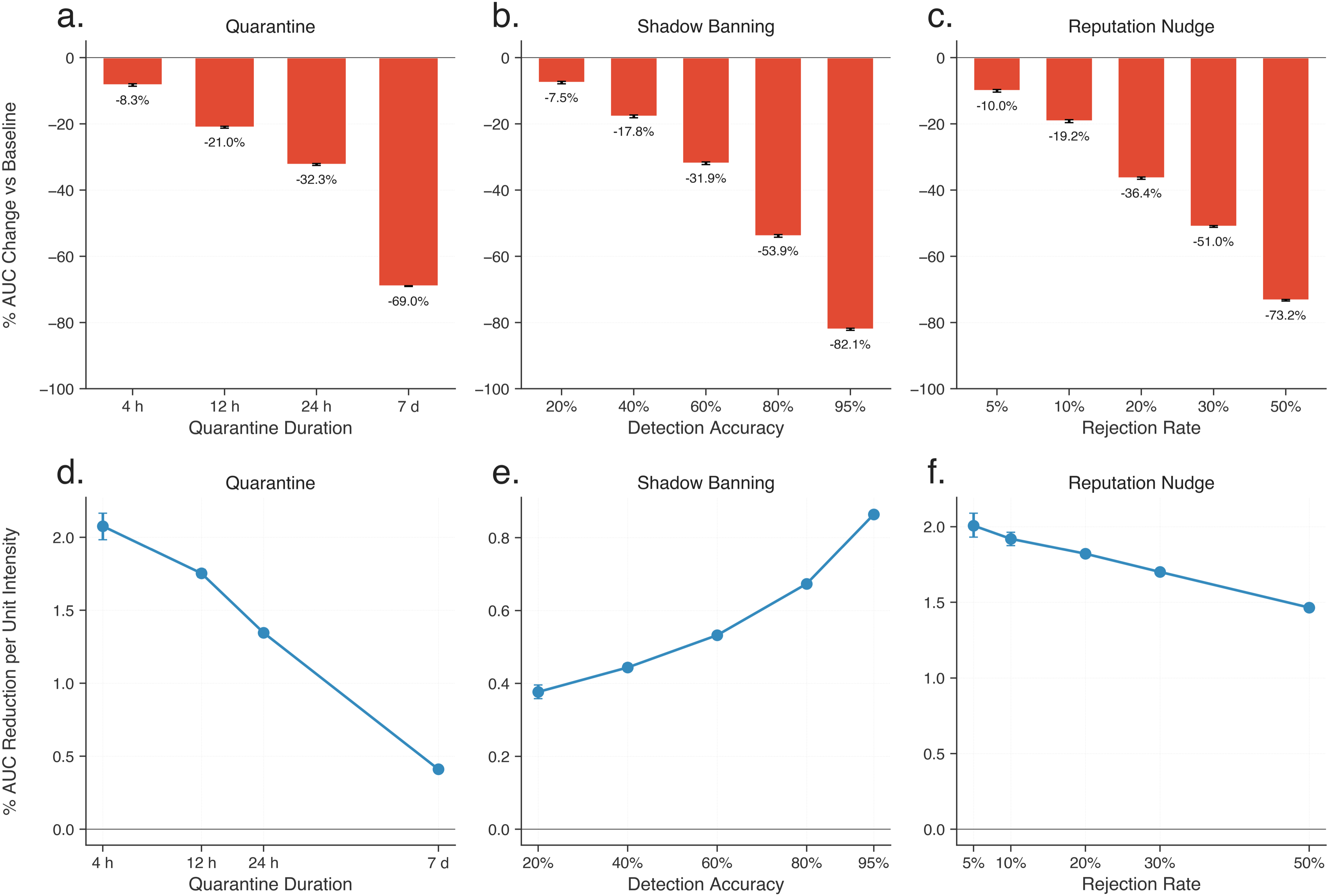}
    \caption{\textbf{Policy counterfactuals across three intervention strategies.} \textit{Top panels (a through c):} total \% AUC reduction relative to the baseline simulation (calibrated to 21{,}683 empirical adoptions; same simulated steps as baseline; error bars: 95\% bootstrap CIs). \textit{Bottom panels (d through f):} per-unit intensity efficiency (\% AUC reduction divided by intervention intensity). \textit{(a, d)} \textit{Quarantine}: post-adoption read-only lockout (4~h, 12~h, 24~h, 1~week); even a one-week lockout achieves only $\sim$69\% reduction, and efficiency falls from $\sim$2.08 at 4~h to $\sim$0.41 at 168~h. \textit{(b, e)} \textit{Shadow banning}: silent suppression of conspiracy posts at detection accuracies 20\% through 95\%; convex dose-response, with $\sim$82\% reduction at 95\% detection. \textit{(c, f)} \textit{Reputation nudge}: warning before first endorsement at rejection rates 5\% through 50\%; rejected adoptions yield permanent conspiracy-specific immunity. The nudge matches shadow banning at a fraction of the intensity because preventing the adoption event simultaneously prevents the state shift that would have facilitated subsequent adoptions.}
    \label{fig:counterfactuals}
\end{figure}

In absolute terms, all three interventions reduce simulated spread, but they do so to different degrees (Fig.~\ref{fig:counterfactuals}a--c). A one-week quarantine reduces AUC by about 69\%, whereas shorter lockouts have substantially weaker effects. Shadow banning can be more effective, but only at high detection rates: 95\% suppression reduces AUC by about 82\%. Reputation nudges reduce AUC by about 73\% at a 50\% rejection rate and achieve reductions comparable to much stronger visibility suppression at lower intensities.

The dose-response curves show why these differences arise (Fig.~\ref{fig:counterfactuals}d--f). Quarantine has diminishing returns as lockout duration increases, because longer lockouts increasingly target the decaying tail of the post-adoption risk window. Shadow banning has a convex response: low levels of suppression leave enough surviving posts to sustain exposure, whereas near-saturation suppression substantially disrupts transmission. Reputation nudges are more efficient at lower intensities because preventing adoption blocks both the outward transmission signal and the within-user susceptibility shift.

These results reflect the interacting-contagion structure of the rabbit hole process. Interventions that delay activity or suppress visibility act after adoption has already changed later susceptibility. Interventions that prevent adoption act earlier, blocking both the outward transmission signal and the within-user shift in susceptibility that would otherwise facilitate later adoption.

\section{Discussion}

This paper set out to explain why conspiracy theories are rarely adopted one at a time. We argued that conspiracy-theory rabbit holes are not a separate mechanism, but the observable trajectory produced when conspiracy theories semantically interact. The results support this account. After users publicly endorse one conspiracy theory, their risk of endorsing another rises; this elevated susceptibility becomes stronger and more persistent with adoption depth; and later adoption is structured by semantic proximity. The results indicate that conspiracy-theory adoption is not simply the repeated expression by high-propensity users, nor a generic slide into any available conspiracy theory. It is a sequential process in which prior adoption changes the conditions under which subsequent conspiracy theories become adoptable.

The findings connect the monological-belief-system tradition in conspiracy research with a dynamic account of social contagion. Prior work has shown that conspiracy beliefs cluster, that individuals differ in their general susceptibility, and that some conspiracy theories often appear as entry points \citep{goertzel1994,wood2012,imhoff2014,brotherton2013,greve2022}. Our results add a dynamic account of how these patterns unfold over time. Public endorsement leaves a measurable trace in the user's subsequent event history: later adoption becomes more likely, the elevated-risk window lengthens with repeated adoption, and transitions concentrate among semantically related conspiracy theories. Stable predispositions may explain why some users adopt more conspiracy theories overall, but they do not by themselves explain why risk rises after adoption, why that risk compounds with depth, or why the next adoption tends to follow semantic structure.

This perspective also reframes gateway conspiracy theories. Gateways are often described as relatively plausible or moderate theories that lead users toward more extreme ones \citep{greve2022}. In an interacting-contagion system, however, gateway status need not stem from a theory’s intrinsic moderation or extremity. Rather, the conspiracy theories that function as gateways should be those closest to many users’ current belief systems. Their high contagiousness follows from this alignment, not from fixed properties of the theory itself.

For contagion theory, the paper highlights a form of interaction that occurs through changing susceptibility within individuals. Standard models of social contagion emphasize transmission between people: one user's sharing behaviour changes another user's probability of adoption. Social channels matter here, since peer exposure predicts adoption and becomes more cumulative after first adoption, consistent with complex contagion accounts of reinforcement \citep{centola2007complex,guilbeault2018complex}. But explaining the rabbit-hole pattern also benefits from characterizing the within-user process as well. The ecology-of-contagions framework theorizes that adopting a conspiracy theory reshapes the user’s belief system toward the adopted conspiracy theory. Under this account, interaction among conspiracy theories arises indirectly: each adoption changes the belief system through which later conspiracy theories are evaluated. This provides a way to study many interacting social contagions without estimating a separate interaction term for every pair of conspiracy theories \citep{guilbeault2018complex,hebert-dufresne2020,hebert-dufresne2025}. 

The semantic results are central to this interpretation. If adoption merely produced a generic state of arousal or attention, then later adoption should rise broadly across conspiracy theories. Instead, adoption follows the geometry of semantic space. Semantically close conspiracy theories are more likely to be co-adopted, transitions are faster within semantic regions than into regions the user has not yet entered, and the settler pattern appears under semantic clustering but not under temporal null clusterings. First entry into a new semantic region is slow, but after entry, later adoption within that region accelerates. This pattern is consistent with prior adoption opening semantic paths rather than simply increasing susceptibility to all conspiracy theories. More broadly, the result is consistent with approaches that represent belief systems in structured relational or latent spaces, rather than as independent attitudes \citep{boutyline2017belief,aiyappa2024emergence,lee2025}. 

The distinction between contagiousness and potency extends this logic. Common entry points are not necessarily the conspiracy theories that most strongly accelerate second adoption. If adoption shifts a user's belief system toward the adopted conspiracy theory, and if larger shifts occur when the adopted theory is farther from the user's current belief-system configuration, then contagiousness and potency should often be in tension. Conspiracy theories close to many users' current positions should be easier to adopt first, making them more contagious, but their adoption may produce limited downstream reshaping. Conspiracy theories farther from users' current positions should be harder to adopt, but once adopted they may induce larger changes and more strongly alter susceptibility to later contagions. This suggests a possible social-contagion analogue of the virulence-transmissibility trade-off in biological pathogens \citep{anderson1982coevolution}. The analogy is structural rather than biological: what makes a contagion easy to spread may differ from what makes it consequential once acquired. The present findings are consistent with this possibility, but the trade-off should be treated as a conjecture for future testing rather than as an established general principle.

The minimal model comparison clarifies what kind of rabbit hole the empirical results document. Independent adoption can generate clustering among users who are generally prone to conspiratorial content, and generic arousal can generate acceleration after entry. Neither reproduces the joint signature of post-adoption acceleration, depth-dependent persistence, semantic concentration and the settler pattern. The ecology-of-contagions model captures this joint signature because adoption increases later susceptibility in a way organized by semantic distance. This comparison is not an experimental proof of private belief change, nor does it rule out all possible alternatives. It shows that the observed temporal and semantic patterns are more consistent with belief-system reshaping than with the specific alternatives tested here.

The simulation counterfactuals illustrate why this ecology of contagions matters for intervention. Recent studies of Community Notes show that notes reduce engagement with a misleading post once attached, especially when they appear early in the post's diffusion \citep{chuai2026community, slaughter2025community}. Yet these estimates focus on the noted post and, in some cases, its diffusion cascade. Our results suggest that such estimates may capture only part of an intervention's value when contagions interact. Preventing adoption of one narrative can also reduce later adoption of related narratives that are never themselves flagged. Early interventions may therefore matter not only because they stop a post before it spreads widely, but because they interrupt cross-contagion spillovers that would otherwise make adjacent narratives more likely to diffuse. Future work should develop causal designs that can estimate these spillovers directly, tracing whether interventions on one misleading post alter subsequent engagement with related, unflagged misinformation.

Several limitations follow from the design. The underlying dataset represents public endorsement on X, not private belief, all content consumption, or all exposure. Some users may privately believe conspiracy theories without publicly endorsing them, and some exposure occurs outside retweets and replies. The analysis is observational, so it does not establish experimental causality or directly observe belief-system change. Depth-dependent hazards could also partly reflect dynamic selection on unobserved frailty, as users with a higher underlying propensity to endorse conspiracy theories may adopt earlier and reach greater adoption depths  \cite{vaupel1985heterogeneity}. This possibility is not inconsistent with the ecology-of-contagions account, since some users’ belief systems may simply be more closely aligned with conspiracy theories from the outset. Such stable differences may help explain why some users adopt more conspiracy theories overall, but they do not by themselves explain why adoption risk changes following prior adoption or the semantic patterns we observe across different conspiracy theories. The endorsement classifiers are validated against human labels and use a conservative threshold, however automated classification can still miss ambiguous cases but that is part of the ambiguity of language and also applies to human labeling. Finally, the setting is the first wave of COVID-19, when conspiracy theories, platform practices and public attention were shaped by a particular crisis. Future work should test whether the same signatures appear on other platforms, during other periods, and for other domains of interacting social contagion.

Taken together, the results suggest that conspiracy-theory rabbit holes are best understood as structured descent through an ecology of contagions. Social exposure brings conspiracy theories to users, but adoption changes what becomes easier to adopt next. Semantic distance shapes that process, making some transitions faster, some regions harder to enter, and some conspiracy theories more consequential after adoption than their first-adoption rates would suggest. The broader implication is that social contagions should not be treated as independent items spreading through a passive network. When adoption reshapes susceptibility, contagions interact through the belief systems of the people who carry them.

\section{Methods}

\subsection*{Data, narrative discovery and endorsement classification}

The dataset comprises 7,616,554 tweets from 7,416 Twitter/X users collected during the first wave of the COVID-19 pandemic, including retweets and replies. Data collection began from two prominent COVID-19 conspiracy hashtags, \textit{\#coronahoax} and \textit{\#virushoax}, streamed between February and May 2020 using the platform's public API. We then retrieved the complete available tweet histories for all identified accounts across the observation window. The seed hashtags define the initial user population, while adoption events are measured from users' broader tweet histories rather than only from tweets containing the seed hashtags.

We exclude likely automated accounts, defined as accounts with Botometer scores above 0.4, as focal subjects in the event-history analyses but remain in the interaction graph and may contribute to social reinforcement for retained users. All timestamps are converted to hourly resolution, measured as continuous hours since the first observed tweet in our dataset.

We identified candidate conspiracy-theory narratives using an embedding-assisted discovery pipeline. Tweets were embedded using a COVID-Twitter-BERT model further tuned on the corpus using SimCSE. We reduced the resulting embeddings using UMAP and clustered them using HDBSCAN to identify semantically coherent regions of discourse. We interpreted these clusters using representative tweets, TF-IDF keywords and domain knowledge. This procedure identified 16 candidate conspiracy-theory narratives, summarized in Table~\ref{tab:categories}.

Because narrative discovery identifies candidate topics rather than endorsement, we validated endorsement directly against human labels. For each narrative, we used keywords from the embedding clusters to sample candidate tweets and hand-labeled an initial set of 500 tweets. When the initial labeled set produced severe class imbalance, we labeled additional positive examples to improve model tuning and validation. Tweets were coded as endorsements only when they explicitly affirmed the relevant conspiracy theory; tweets that merely mentioned, criticized or discussed adjacent misinformation were coded as non-endorsements.

These human labels formed the basis for model development and validation. For each narrative, we split the labeled tweets into training and held-out test sets and trained a separate prompt-tuned BLOOMZ-1.7B classifier. The classifiers were evaluated against the held-out human labels, ensuring that the measurement procedure approximated our human coding rule rather than simply detecting conspiracy-related vocabulary. In the main analyses, a tweet is classified as endorsing a conspiracy theory only when the corresponding classifier assigns a positive endorsement probability of at least 0.80. This conservative threshold prioritizes precision over recall to avoid false positives. Full prompt-tuning specifications, model diagnostics and error analyses are reported in the Supplementary Information.

\subsection*{Adoption histories}

The classification pipeline identifies 16 conspiracy-theory narratives. For the empirical analyses, we merge the Anthony Fauci and Hydroxychloroquine narratives because they occupy highly overlapping regions of semantic space and are substantively intertwined in the endorsement data. This yields 15 analytic conspiracy theories.

We define adoption as the first observed time on which a user publicly endorses a conspiracy theory in a tweet. Later endorsements of the same theory are treated as repeat sharing rather than additional adoption events. For each user, we reconstruct a temporal adoption history recording the sequence and timing of conspiracy-theory adoption. Adoption depth is defined as the number of distinct conspiracy theories a user had previously adopted at a given point in time.

\subsection*{Semantic distance and semantic regions}

We measure semantic proximity among the 15 analytic conspiracy theories using the same embedding space used in the narrative-discovery pipeline. For each conspiracy theory, we collect the embeddings of tweets classified as endorsements of that theory. We then calculate the semantic distance between each pair of conspiracy theories as the mean pairwise Euclidean distance between their endorsement-tweet embeddings. Smaller values indicate that two conspiracy theories occupy more similar regions of the embedding space. For the merged Fauci/Hydroxychloroquine theory, the endorsement set combines tweets classified as endorsing either of the two original narratives. This procedure yields a 15-by-15 semantic distance matrix.

We use this distance matrix to identify broader semantic regions among conspiracy theories. This clustering step is distinct from the earlier HDBSCAN clustering used for narrative discovery: the earlier procedure clusters individual tweets to identify candidate narratives, whereas this procedure clusters already-identified conspiracy theories. We apply hierarchical clustering to the 15-by-15 semantic distance matrix and select the number of clusters using silhouette score. This yields 6 semantic regions which are used in the within-cluster transition analyses, the settler-pattern analyses and the cross-cluster covariate in the sequential hazard models. To assess whether these patterns reflect semantic structure rather than temporal availability, we also construct temporal null clusters based on first-appearance timing and peak-frequency timing of each conspiracy theory.

\subsection*{Network reconstruction and peer exposure}

We construct an undirected user interaction network based on retweet and reply ties. An edge connects two users whenever one user retweets or replies to the other. For the hazard analyses, we remove isolated users from the network.

\subsection*{Sequential hazard models}
We modeled adoption as an ordered sequence of distinct conspiracy narratives. For each user and narrative, adoption time was defined as the first hour in which the user shared a tweet classified as that narrative. Models were fit among focal users classified as human using a Botometer threshold of less than or equal to 0.4. Model 1 estimated entry into the adoption sequence, defined as time to first narrative adoption. Model \(k\) estimated the time until a user adopted their \(k\)-th distinct narrative, conditional on having already adopted \(k-1\) narratives. At each step, the risk set contained narratives that the user had not yet adopted and that had already appeared in the data.
Risk intervals used delayed entry and right censoring. In Model 1, users entered risk at the later of their first observed activity and the first appearance of the candidate narrative. In later models, users entered at the later of their previous adoption time and the first appearance of the candidate narrative. Users who experienced the relevant next adoption contributed an event for the adopted narrative and censored observations for the remaining candidate narratives at the same exit time. Users who did not experience the relevant next adoption were censored one hour after their last observed activity.
We included peer exposure as a time varying covariate, measured every 8 hours as the number of unique graph neighbors who had shared the candidate narrative during the previous 14 days. Exposure entered the model as indicators for one, two, three, and four or more exposed neighbors. Models also included log degree and narrative indicators, with Fake News as the reference narrative.
For adoption after entry into the sequence, the event clock was reset at the user’s most recent prior adoption. Model 2 was estimated in two specifications. Model 2a controlled for the candidate second narrative. Model 2b replaced these candidate narrative indicators with indicators for the narrative adopted first, allowing us to estimate which first adoptions most strongly accelerated adoption of another narrative next. We define contagiousness using the narrative specific hazard ratios from Model 1 and potency using the first adoption coefficients from Model 2b. Models 2 through 8 also included an indicator for whether the candidate narrative belonged to a semantic region not previously visited by the user.
Simultaneous adoption times were placed in a randomized but reproducible order and separated by one minute offsets, or by smaller equally spaced offsets when one minute spacing would have overtaken the next observed event. To obtain smooth stage specific baseline hazards for comparison and simulation, we parameterized each model’s estimated cumulative baseline hazard. We fit a through origin linear function for Model 1 and two parameter Weibull functions for Models 2 through 8, then differentiated the fitted curves to obtain instantaneous hazards. These hazards were used for the comparisons at 1, 24, and 72 hours, the recovery window calculations, and the agent based simulation, while retaining each model’s centered covariate reference profile. We fit models through the eighth adoption step, stopping before Model 9 because it fell below the prespecified 1000 event minimum.

\subsection*{Minimal model comparison}

The main text compares three minimal models: the \textit{No Interaction Model}, the \textit{General Arousal Model}, and the \textit{Ecology of Contagions Model}. Here we give the formal specification. The models are used for mechanism discrimination and visualization rather than for estimating private belief states.

All three models share the same baseline adoption process. Users' belief systems and conspiracy narratives are represented in a shared latent belief-system space. User $i$'s belief-system state is represented by $\theta_i(t) \in \mathbb{R}^D$, and narrative $c$ has fixed position $\sigma_c \in \mathbb{R}^D$ in the same space. The adoption hazard is
\begin{equation}
h_{ic}(t)
=
\mu A_i(t)
\exp\left(
-\frac{\|\sigma_c - \theta_i(t)\|^2}{\rho}
\right),
\end{equation}
where $\mu$ is the base adoption rate, $\rho$ controls resistance to distance in the latent belief-system space, and $A_i(t)$ captures non-specific post-adoption arousal. The distance term is common to all three models: narratives closer to the user's current belief-system state are more readily
adopted than narratives farther away.

The models differ only in the post-adoption terms governed by $\alpha$ and $\beta$. When user $i$ adopts narrative $c$ at time $t_a$, the belief-system state may shift toward the adopted narrative:
\begin{equation}
\theta_i(t_a^+)
=
(1-\alpha)\theta_i(t_a^-)
+
\alpha\sigma_c.
\end{equation}
Thus, $\alpha=0$ leaves the belief-system state unchanged, while $\alpha>0$ implements belief-system reshaping. In the absence of further adoption, the belief-system state relaxes toward baseline $\theta_i^0$:
\begin{equation}
\frac{d\theta_i}{dt}
=
k(\theta_i^0 - \theta_i),
\end{equation}
where $k$ controls the rate of return toward baseline. 
The general-arousal term resets after adoption and decays exponentially:
\begin{equation}
A_i(t)
=
1 + \beta \exp[-\gamma(t-t_a)],
\end{equation}
with $A_i(t)=1$ before first adoption. Thus, $\beta=0$ implies no generic post-adoption arousal, while $\beta>0$ temporarily raises susceptibility to all contagions uniformly.

\begin{table}[H]
\centering
\caption{The three minimal models. All models share the same distance-based
baseline adoption hazard and differ only in their post-adoption mechanism.}
\label{tab:variants}
\begin{tabular}{lccp{6cm}}
\toprule
\textbf{Model} & $\boldsymbol{\alpha}$ & $\boldsymbol{\beta}$ & \textbf{Post-adoption mechanism} \\
\midrule
\textit{No Interaction Model} & $0$ & $0$ &
No post-adoption change in belief-system state or general susceptibility. \\
\textit{General Arousal Model} & $0$ & $>0$ &
Temporary, content-neutral increase in susceptibility after adoption. \\
\textit{Ecology of Contagions Model} & $>0$ & $0$ &
Belief-system reshaping toward the adopted narrative. \\
\bottomrule
\end{tabular}
\end{table}

\subsection*{Bootstrapping and statistical tests}
Point estimates were calculated from the full sample. For the Cox models, we obtained 95 percent percentile confidence intervals from 200 user level bootstrap samples. Users were sampled with replacement from the master population of 4,772 users, and all counting process rows belonging to each sampled user were retained. Repeatedly sampled users received unique identifiers. We refitted all eleven model specifications and recomputed coefficient hazard ratios, peer exposure contrasts, gateway coefficients, parametric baseline hazards, temporal hazard ratios, and recovery windows within each sample. Confidence limits were the 2.5th and 97.5th percentiles of the bootstrap distributions. Cox and gateway intervals were pointwise, without multiplicity adjustment across narrative specific coefficients.

For semantic transition summaries, we used 2,000 bootstrap samples of complete user adoption timelines. Resampling preserved each user’s full sequence, while the selected semantic, first appearance, and peak frequency cluster assignments remained fixed. Within each sample, we recomputed median waiting times for transitions within previously entered clusters, jumps into previously unvisited clusters, and the three phases of settler sequences. These intervals capture sampling variation among users but not uncertainty in cluster selection.

We tested the association between semantic distance and one minus Jaccard coadoption similarity using a one sided Mantel test. The test statistic was the Spearman correlation between the upper triangular matrix entries. Significance was assessed using 9,999 joint row and column permutations, with a prespecified positive alternative and the plus one correction. Within cluster and jump transition times were compared using a one sided Mann Whitney U test. Settler sequences were evaluated using a Friedman test followed by two sided Wilcoxon signed rank tests, with Holm correction across the three pairwise comparisons within each clustering family. Users could contribute multiple transitions or settler sequences, so these rank tests treated transitions or matched triads as the analysis units.

For policy counterfactuals, we matched the 100 intervention and baseline runs by run identifier and resampled these pairs 2,000 times. Each sample was used to recompute the percentage reduction in mean total AUC. The resulting intervals describe stochastic variation across simulation runs and do not propagate uncertainty in the fitted Cox or Hawkes parameters. Hawkes and Poisson models were compared using AIC, while simulation fidelity statistics were treated as descriptive.

\subsection*{Agent based network simulation}

We embedded the fitted event history dynamics in an agent based simulation on the observed human retweet and reply network. The simulation used the largest connected component after removing accounts classified as bots, yielding 4,516 users and 20,808 undirected interaction edges. The simulation included the 15 analytic conspiracy theories. Each agent retained its network degree, adoption history, semantic region history and sharing history.

The simulation proceeded in hourly steps. In the baseline configuration, all contagions entered the simulation at hour 1. For each contagion, one percent of eligible users was sampled at random and assigned an initial seed adoption. These seed adoptions created the initial visible shares from which network exposure could grow. Seeds were included in adoption and sharing histories, but subsequent adoption was generated by the fitted sequential Cox hazard models rather than by hand tuned diffusion rules.

At each hour, each active user faced a competing risk over contagions they had not yet adopted. The hazard model was selected from the user's current adoption depth: first adoption used the first adoption model, later adoptions used the corresponding sequential models through Model 8, and deeper adoption histories reused the deepest fitted model. The simulated hazards retained the fitted baseline hazard, peer exposure dose response, log degree, target contagion indicators and semantic region transition term. Peer exposure was recomputed at every hour as a time varying covariate, counting unique neighbours who had visibly shared the candidate contagion during the previous 14 days. Candidate conspiracy theories with no visible neighbour exposure were not eligible for organic adoption. Candidate hazards were summed to obtain the probability of any adoption in that hour, and the adopted contagion was drawn in proportion to its hazard, allowing at most one new adoption per user per hour.

Adoption also recorded the user's first share of the adopted contagion. Repeat posts after adoption were generated by a pooled Hawkes self exciting process fitted to empirical user by contagion sharing sequences. The Hawkes process included a baseline repeat sharing rate and exponential self excitation from the user's own prior posts. Repeat posts were drawn from the integrated hourly Hawkes intensity and visible repeat posts contributed to future peer exposure for neighbours.

Because the simulation does not include exogenous news shocks, platform events or other forces that determine the end of the empirical observation window, baseline runs were stopped after the first hourly update in which the cumulative number of distinct user by contagion adoptions reached or exceeded the empirical target of 21,683. This stopping rule calibrates total adoption volume to the empirical target while allowing the timing, ordering and composition of adoption to arise endogenously. Counterfactual scenarios were then run for the mean baseline stopping time, 2,181 hours, so that intervention effects reflect changes in diffusion over a common time horizon. We ran 100 independent replications per scenario, using the run identifier as the random seed.

\subsection*{Policy counterfactuals and uncertainty}

We evaluated three stylized policy counterfactuals that intervene at different points in the interacting contagion process. All counterfactuals used the same empirical network, seeding procedure, fitted sequential hazard structure and Hawkes repeat posting process as the baseline agent based simulation.  Counterfactual runs were evaluated over the mean baseline duration of 2,181 simulated hours, with 100 independent replications per scenario.

\textit{Quarantine} imposes a temporary read only period after an organic adoption, where organic denotes an adoption generated by the sequential hazard model rather than a seed. During this period, the user cannot adopt additional contagions and cannot generate Hawkes repeat posts for any contagion. Existing visible posts remain visible to neighbours and continue to contribute to exposure. We evaluated lockout durations of 4, 12, 24 and 168 hours (7 days). Seed adoptions do not start the quarantine clock.

\textit{Shadow banning} implements silent visibility suppression. Each organic adoption share and each Hawkes repeat post is hidden from neighbours with probability equal to the detection rate. Hidden posts do not enter neighbours' 14 day exposure histories, but they remain in the poster's own sharing history and can still contribute to that user's Hawkes self excitation. Initial seed shares are exempt. We evaluated detection rates of 20, 40, 60, 80 and 95 percent.

\textit{Reputation nudges} act before public endorsement. When the competing risk draw selects an organic adoption, the adoption is rejected with probability equal to the nudge rate. A rejected adoption is replaced by permanent immunity for that user and conspiracy theory, so the user cannot later adopt the same conspiracy theory, while other conspiracy theories remain eligible. We evaluated rejection rates of 5, 10, 20, 30 and 50 percent.

For each run and contagion, we computed the adoption fraction at each simulated hour and integrated this curve over time. Total AUC is the sum of these contagion specific AUCs within a run. Intervention effects are reported as percentage reductions in total AUC relative to the baseline. For uncertainty, we matched intervention runs to baseline runs by run identifier and resampled these matched run pairs with replacement 2,000 times. In each resample, we recomputed the mean scenario to baseline AUC ratio. Intervals are the 2.5 and 97.5 percentiles of the resulting bootstrap distribution. These intervals reflect stochastic simulation variability and do not propagate uncertainty from the fitted hazard model or Hawkes process.

\subsection*{Data and code availability}
The public replication package, including the pseudonymized data and code needed to reproduce the reported analyses, simulations, formal model results, tables and manuscript figures, is available at \url{https://github.com/ftschofenig/rabbit-hole-interacting-contagions}. To protect user privacy, the released data contain Tweet identifiers rather than Tweet text and a pseudonymized interaction network with the timestamps of the sharing times of each and every conspiracy theory without usernames, handles, profile information or URLs. Researchers may rehydrate the Tweet identifiers where content remains available, subject to platform access and terms.

\section{Acknowledgements}
We are grateful for funding support from the Business, Government, and Society (BGS) Initiative at the Stanford Graduate School of Business.

\section{Competing interests}
The authors declare no competing interests.

\newpage
\clearpage

\bibliographystyle{apalike}
\bibliography{references}

\newpage

\appendix

\setcounter{section}{0}
\renewcommand{\thesection}{S\arabic{section}}

\setcounter{subsection}{0}
\renewcommand{\thesubsection}{\thesection.\arabic{subsection}}

\setcounter{figure}{0}
\renewcommand{\thefigure}{S\arabic{figure}}

\setcounter{table}{0}
\renewcommand{\thetable}{S\arabic{table}}

\setcounter{equation}{0}
\renewcommand{\theequation}{S\arabic{equation}}

\section*{Supplementary Information}
\addcontentsline{toc}{section}{Supplementary Information}

\section{Data processing pipeline}\label{apx:sec_data_processing_pipeline}
\subsection*{Measuring Conspiracy Endorsement}
The following section outlines our process of transforming a corpus of tweets into a user-level temporal sequence of conspiracy endorsements. This goal consists of two distinct tasks: 1) determining the set of possible conspiracy theories in the data, and 2) classifying tweets as endorsing (or not endorsing) a given conspiracy theory.
\subsubsection*{Tweet Embeddings}
We represent the semantic landscape of conspiracy theories by embedding each conspiracy tweet into a high dimensional embedding space. Of course, the efficacy of this approach hinges upon the ability to generate high-quality tweet embeddings, and LLMs produce notoriously poor-quality representational embeddings \citep{li2020sentence,rajaee2021cluster}. We take two steps to significantly improve the quality of embeddings. First, we use COVID-Twitter-BERT, a BERT-large-uncased model trained on 22.5 million tweets from January to April 2020, as our base model to capture both the syntactic style of Twitter and semantics relating to COVID-19. Second, we finetune the BERT model on our own corpus of tweets using SimCSE (Simple Contrastive Learning of Sentence Embeddings) \citep{gao2021simcse}. This deceptively simple method takes an input, in our case a tweet, and predicts itself using standard dropout as noise. By reducing anisotropy and increasing uniformity in the embedding space, SimCSE produces more semantically meaningful document embeddings. We finetune our model by feeding the {[}CLS{]} token from the last layer BERT layer in SimCSE for one epoch and generate 1024-dimension tweet embeddings as the last layer {[}CLS{]} token from the fine-tuned model.

\begin{figure}
    \centering
    \includegraphics[width=0.75\linewidth]{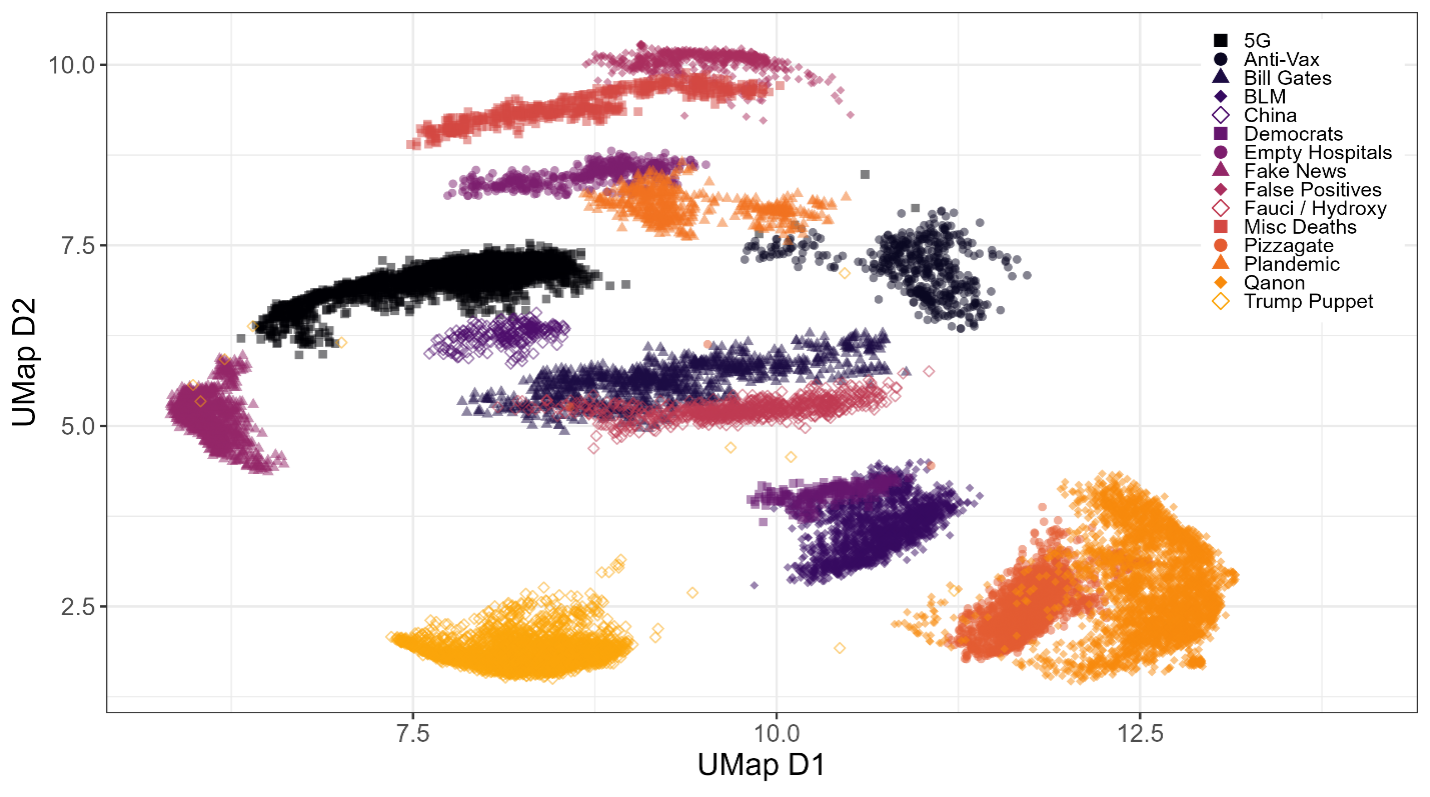}
    \caption{Two-dimensional visualization of conspiracy tweet topics. Clusters are produced via 10-dimensional U-MAP dimensionality reduction and clustered via HDBSCAN. For full details on figure construction please refer to the Appendix text.}
    \label{fig:UMAP_visualization}
\end{figure}

We reduce the dimensionality of the embedding space to 10 dimensions using uniform manifold approximation and projection (UMAP) to make clustering computationally feasible and identify clusters in this space using Hierarchical Density-Based Spatial Clustering of Applications with Noise (HDBSCAN) \citep{gao2021simcse, McInnes2018}. We identify the keywords associated with each cluster by applying TF-IDF to each cluster. Hyperparameters for both algorithms were jointly tuned to maximize the interpretability and distinctiveness of conspiracy clusters (Final hyperparameters: UMAP: number of neighbors=300; minimum distance=0; cosine similarity; HDBSCAN: minimum samples=2; minimum cluster size=100, cluster selection epsilon=0.01; Euclidean distance). These clusters and their keywords can be seen as analogous to topic models, where each document is assigned to only a single topic. Table \ref{tab:categories} lists the 16 conspiracy theories identified in the data using this approach and a two dimensional visualization is displayed in Figure~\ref{fig:UMAP_visualization}.

\subsubsection*{Prompt-tuned LLM Classification}
To build a labeled training set for each conspiracy, we use the signifying keywords from the embedding clusters to identify a set of tweets that may promote the conspiracy theory. For conspiracy theories involving Bill Gates, for instance, tweets were filtered on regular expression matches for the strings \emph{bill} and \emph{gates}, and the research team hand-labeled 500 tweets. Importantly, our theoretical interest is in the endorsement of conspiracy theories, not misinformation more broadly. As a result, coders labeled tweets that contained misinformation but did not explicitly endorse a conspiracy as 0 (no endorsement). For instance, there were many tweets suggesting that Hydroxychloroquine was an effective treatment for COVID-19. These tweets were not labeled as conspiratorial unless they suggested some nefarious action as well, like the CDC suppressed evidence of the drug's efficacy. When a conspiracy had severe class imbalance, often due to too many negative training examples, further positive training examples were labeled in the data.

Using these hand-annotated data, we split the sample into an 80/20 train/test split and tuned the prompts using the BLOOMZ 1.7B parameter auto-regressive LLM \citep{muennighoff2023crosslingual}. BLOOMZ extends BLOOM (BigScience Large Open-science Open-access Multilingual Language Model) by finetuning this base model on English natural language prompts and tasks. In addition to the flagship 176B parameter model, the authors have released a series of smaller models, including 560M, 1.1B, 1.7B, 3B, and 7.1B. We use the BLOOMZ 1.7B as we found no meaningful improvements in performance using the 3B and 7.1B models during testing, and 176B was too large for practical utility.

\paragraph{Prompt Initialization.} The time needed to learn the optimal prompt \emph{P} can be significantly reduced if, instead of random initialization, researchers initialize \emph{P} using a natural language prompt. We initialize with a standard few-shot prompt, providing an example positive case, common conspiracy hashtags, and appending ``Answer:''.

\begin{quote}
\textquotesingle\textquotesingle\textquotesingle Do the following tweets endorse the conspiracy theory that China created and spread the covid-19 virus? Common conspiracy hashtags include \#chinaliedpeopledied and \#chinavirus. Answer Yes or No.

Tweet: China deliberately created and spread \#wuhanvirus to cripple the us economy with Fauci funding

Answer: Yes

Tweet: \{TWEET TO CLASSIFY\}\\
Answer:
\end{quote}

When prompt tuning, researchers must also select the length of the synthetic prompt. Longer prompts provide more information for the frozen language model but come at additional computational costs. We search the hyperparameter space using grid search over a set of learning rates (3e-2 to 8e-5), batch sizes (1 to 64), and prompt lengths (8 to 512) and use a final learning rate of 3e-3, batch size of 2, and prompt length of 256. We train the model using a linear decay learning rate and AdamW optimization and end training via early stopping on the evaluation set to avoid overfitting (approximately 4 epochs). With a prompt length of 256 we only learn 524,288 trainable parameters (\textasciitilde0.03\% of 1.7B). Although traditional LLM pipelines eschew text preprocessing, we found that lowercasing the tweet significantly improved model performance. We also found these hyperparameters maximized model performance regardless of the conspiracy theory. We make all our models and code used to prompt tune publicly available and summarize model performance in Table \ref{tab:s1_prompt_tuning_model_performance}. Due to the small model size, any scholar interested in our method can easily run our code for free on Google Colab, with no need for prior deep learning experience.

\begin{table}
\caption{Prompt tuned LLM performance evaluated at the fixed endorsement probability threshold of 0.8, chosen to reduce false positives against human labeling. Accuracy, precision, recall, and F1 compare the thresholded LLM classifications against those human labels.}
\label{tab:s1_prompt_tuning_model_performance}
\begin{tabular}{lllllll}
\toprule
Conspiracy & Count & Human Label Rows & Accuracy & Precision & Recall & F1 \\
\midrule
5G & 11,879 & 500 & 0.93 & 0.97 & 0.90 & 0.93 \\
Anti-Vaccine & 43,974 & 498 & 0.81 & 0.96 & 0.67 & 0.79 \\
Bill Gates & 29,391 & 500 & 0.83 & 0.96 & 0.79 & 0.87 \\
Black Lives Matter & 95,808 & 499 & 0.82 & 0.98 & 0.37 & 0.54 \\
China Created Covid & 78,351 & 751 & 0.90 & 0.98 & 0.54 & 0.69 \\
Democrats & 161,644 & 550 & 0.83 & 1.00 & 0.36 & 0.53 \\
Empty Hospitals & 34,457 & 554 & 0.98 & 0.97 & 0.94 & 0.96 \\
Fake News & 162,361 & 502 & 0.86 & 0.88 & 0.65 & 0.75 \\
False Positive Testing & 126,908 & 578 & 0.92 & 0.87 & 0.73 & 0.79 \\
Anthony Fauci & 30,753 & 500 & 0.92 & 1.00 & 0.77 & 0.87 \\
Hydroxychloroquine & 19,406 & 551 & 0.89 & 0.87 & 0.74 & 0.80 \\
Miscounted Deaths & 137,293 & 592 & 0.94 & 0.91 & 0.85 & 0.88 \\
Pizzagate & 14,109 & 500 & 0.63 & 0.87 & 0.60 & 0.71 \\
Plandemic & 87,772 & 502 & 0.89 & 0.95 & 0.65 & 0.77 \\
QAnon & 56,908 & 501 & 0.69 & 0.96 & 0.53 & 0.68 \\
Trump Puppet & 18,949 & 561 & 0.89 & 0.85 & 0.76 & 0.80 \\
\bottomrule
\end{tabular}
\end{table}

\paragraph{Benchmarking Performance.} Average accuracy and F1 are at par or exceed existing transformer-based conspiracy tweet classification methods, despite using orders of magnitude less training examples and compute time \citep{pogorelov2021wico}. We also benchmark the performance of our prompt-tuned models against a few-shot-prompted 175B parameter GPT-3 Davinci-003 model. Davinci-003 finetunes the baseline GPT-3 model through reinforcement learning from human feedback, meaning the model learns from human labelers how to follow instructions and respond to questions. The few-shot prompt provided to BLOOMZ during training initialization is also provided to GPT-3 to obtain comparable results. We present these results for the Bill Gates conspiracy in Table \ref{tab:s2_prompt_tuning_benchmark}, and find our approach outperforms the baseline few-shot GPT-3 model by roughly 22 percentage-points in accuracy and F1.

\begin{table}[htbp]
\centering
\caption{Prompt Tuning Benchmark}
\label{tab:s2_prompt_tuning_benchmark}
\begin{tabular}{lrr}
\toprule
 & BLOOMZ-1.7B Prompt Tuned & GPT-3 175B Davinci-003 \\
\midrule
Accuracy & 0.85 & 0.63 \\
F1 & 0.904 & 0.697 \\
Precision & 0.887 & 0.52 \\
Recall & 0.903 & 0.929 \\
\bottomrule
\end{tabular}
\end{table}

\paragraph{Error Analysis.} One particularly useful feature of language model prediction is that we can obtain predicted probabilities for the classification task by applying a normalized exponential function to the log probabilities of token estimation. This can be particularly useful to avoid false positives, where we inadvertently flag a user for propagating a conspiracy theory when they did not. Consistent with this framework, we filter the test set to predictions where the model outputs a token likelihood of 80 percent or greater, indicating strong model confidence, and accuracy and F1 jump to 0.92 and 0.95 on this subsample, respectively.

As an example, we examine the three instances in which the model was confidently incorrect (incorrect token likelihood \textgreater90\%), presented in Table \ref{tab:s3_model_misclassification}. Although our labelers rate the first two examples as non-conspiratorial, a detailed review reveals that the tweets do in fact endorse conspiratorial content. The first tweet suggests Bill Gate\textquotesingle s involvement in a separate conspiracy theory that the Pirbright Institute patented the COVID-19 genome and deliberately released it as a biological agent. The second tweet promulgates the theory that Bill Gates is pushing climate engineering for his own profit. The final tweet clearly promotes a conspiracy theory--the QAnon conspiracy--but makes no reference to Bill Gates. It was sampled for analysis for using the term ``gates'' and is a true classification error. On the whole, however, this error analysis suggests the model is comparable to, or perhaps even outperforms, human judgement, with a true precision exceeding 0.98 on this small sample.

\begin{table}[htbp]
\centering
\caption{Model Misclassification}
\label{tab:s3_model_misclassification}
\small
\begin{tabularx}{\textwidth}{>{\raggedright\arraybackslash}Xccc}
\toprule
 & Human Label & Model Label & Model ``Yes'' Likelihood \\
\midrule
European coronavirus issued in November 2019 to the Pirbright Institute from Surrey, funded by the Gates
& No & Yes & 0.912 \\

Billll Gates holds at least one patent for Climate Engineering technology. He also has links/investments (\$00m) in Sourh American avaiarion Co. developing / testing the technologies associated with Climate Engineering. Need for CE based on flawed assumptions the world is warming.
& No & Yes & 0.975 \\

I’ve seen the pic on website this morning, if no one corroborates this, Im in London 2morra and I’ll go down and take a photo of the gates myself to prove if it’s a fake I hope its not \#WWG1WGAWORLDWIDE \#Qanons \#TrustThePlan \#DarkToLight \#TheGreatAwakening \#SaveTheChildren
& No & Yes & 0.910 \\
\bottomrule
\end{tabularx}
\end{table}

\subsubsection*{Classifying Tweet-Level Endorsement}

The ultimate goal of this machinery is to determine whether a given tweet endorses a conspiracy, \emph{C}. As we are particularly concerned with false positives, we use a conservative threshold when determining endorsement. Formally, we say that tweet \emph{X} endorses conspiracy \emph{C} if the predicted probability of positive classification is at least 80\%.

\[
{Endorsement}_{C}\left( X \right) =
\left\{
\begin{aligned}
1,\ \  & \text{if } \Pr_{\theta}(Y = \text{`Yes'} \mid \theta_{P};X) \ge 0.8 \\
0,\ \  & \text{otherwise}
\end{aligned}
\right.
\]

We retain only tweets classified as endorsing a conspiracy under this threshold and drop all other tweets from the analysis. This conservative threshold prioritizes precision over recall and reduces the risk of incorrectly classifying non-endorsement as endorsement.

\section{Hierarchical clustering of conspiracy theories}\label{apx:sec_hierarchical_clustering}

The semantic analyses require a grouping of conspiracy theories into broader regions of the belief system space. We use hierarchical clustering for this purpose, but apply it at the level of already identified conspiracy theories rather than at the level of individual tweets. This step is therefore distinct from the HDBSCAN procedure used earlier for narrative discovery. HDBSCAN identifies candidate conspiracy theory narratives from tweet embeddings; the hierarchical clustering described here groups the resulting analytic conspiracy theories into regions used for transition analyses, settler pattern analyses, and the semantic region indicator in the sequential hazard models.

For the primary semantic clustering, we begin with the \(15 \times 15\) semantic distance matrix among the analytic conspiracy theories. Each entry gives the mean pairwise Euclidean distance between endorsement tweet embeddings for two conspiracy theories. Smaller values indicate that the two conspiracy theories occupy more similar positions in the embedding space. The Fauci/Hydroxychloroquine category uses the combined endorsement set for the two original narratives, matching the merged analytic category used throughout the event history models.

We then apply hierarchical clustering to this distance matrix. In the analysis notebooks, we evaluated Ward, average, complete, and single linkage, with the number of clusters \(k\) ranging from 3 to 8. For each linkage and \(k\), the resulting labels were evaluated using the silhouette score computed on the precomputed distance matrix. The selected solution is the linkage and cluster count with the highest silhouette score. This procedure selected Ward linkage with \(k=6\) semantic regions, with a silhouette score of 0.32. The six semantic regions are: Black Lives Matter and Democrats; QAnon and Pizzagate; Trump Puppet; Deaths, Plandemic, Hospitals, and Testing; Anti Vax, Gates, and Fauci/Hydroxychloroquine; and Fake News, 5G, and China.

To test whether the transition patterns reflect semantic organization rather than simple temporal availability, we construct two temporal null clusterings using the same hierarchical clustering procedure. These null clusterings preserve the fact that conspiracy theories emerged and peaked at different times, but replace semantic distance with temporal distance. The first null clustering is based on first appearance. For each conspiracy theory, we identify the first hour at which it appears anywhere in the analytic graph. The distance between two conspiracy theories is the absolute difference between their first appearance hours. Applying the same linkage and silhouette search selected Ward linkage with \(k=3\), with a silhouette score of 0.69. This clustering grouped Deaths, Gates, Fauci/Hydroxychloroquine, and Testing; Plandemic and China; and Fake News, 5G, QAnon, Anti Vax, Hospitals, Black Lives Matter, Pizzagate, Trump Puppet, and Democrats.

The second temporal null clustering is based on peak sharing frequency. For each conspiracy theory, all sharing events are aggregated across users into an hourly time series. We then compute a centered 24 hour rolling sum and define the peak time as the hour with the largest rolling count. The distance between two conspiracy theories is the absolute difference between their peak sharing hours. The same hierarchical clustering search selected Ward linkage with \(k=3\), with a silhouette score of 0.77. This clustering grouped 5G, Hospitals, and China; Fake News, Deaths, QAnon, Anti Vax, Plandemic, Fauci/Hydroxychloroquine, Testing, Pizzagate, and Democrats; and Gates, Black Lives Matter, and Trump Puppet.

For all three cluster families, transition statistics are computed from human user adoption timelines. For each retained user, we record the first observed adoption time for each conspiracy theory and order these adoptions by time. A within region transition is a consecutive move between two conspiracy theories assigned to the same cluster. A first entry into a new region is a consecutive move in which the next conspiracy theory belongs to a cluster that the user has not previously visited. The settler pattern is measured with four adoption sequences of the form \(A,A,B,B\): one transition within an already visited region, followed by a move into a new region, followed by another transition within that new region. By default, the new region must not have appeared earlier in the user timeline.

The same definitions are applied to the semantic regions, the first appearance clusters, and the peak frequency clusters. Thus, the comparison isolates what changes when the grouping of conspiracy theories is semantic rather than temporal. Under the semantic clustering, users move faster within already entered semantic regions than into new semantic regions, and the settler pattern shows a slow entry into a new region followed by faster adoption within that region. In matched four event settler sequences, median transition times before entry, during entry, and after settling were 21.9, 75.8, and 23.4 hours under semantic clustering. The corresponding medians were 52.5, 61.9, and 62.0 hours under first appearance clustering, and 77.2, 72.3, and 113.6 hours under peak frequency clustering. Thus, only semantic clustering shows a pronounced slowdown during entry followed by faster transitions after settling. Confidence intervals for these medians are computed by resampling full user timelines with replacement and recomputing the median transition statistics in each bootstrap draw.

These temporal null comparisons show that the main result is not simply a consequence of conspiracy theories becoming available at similar points in the observation window or peaking during the same attention cycle. 

\section{Formal model initialization and simulation}\label{sec:formal_model_appendix}
\begin{figure}[H]
    \centering
    \includegraphics[width=0.8\textwidth]{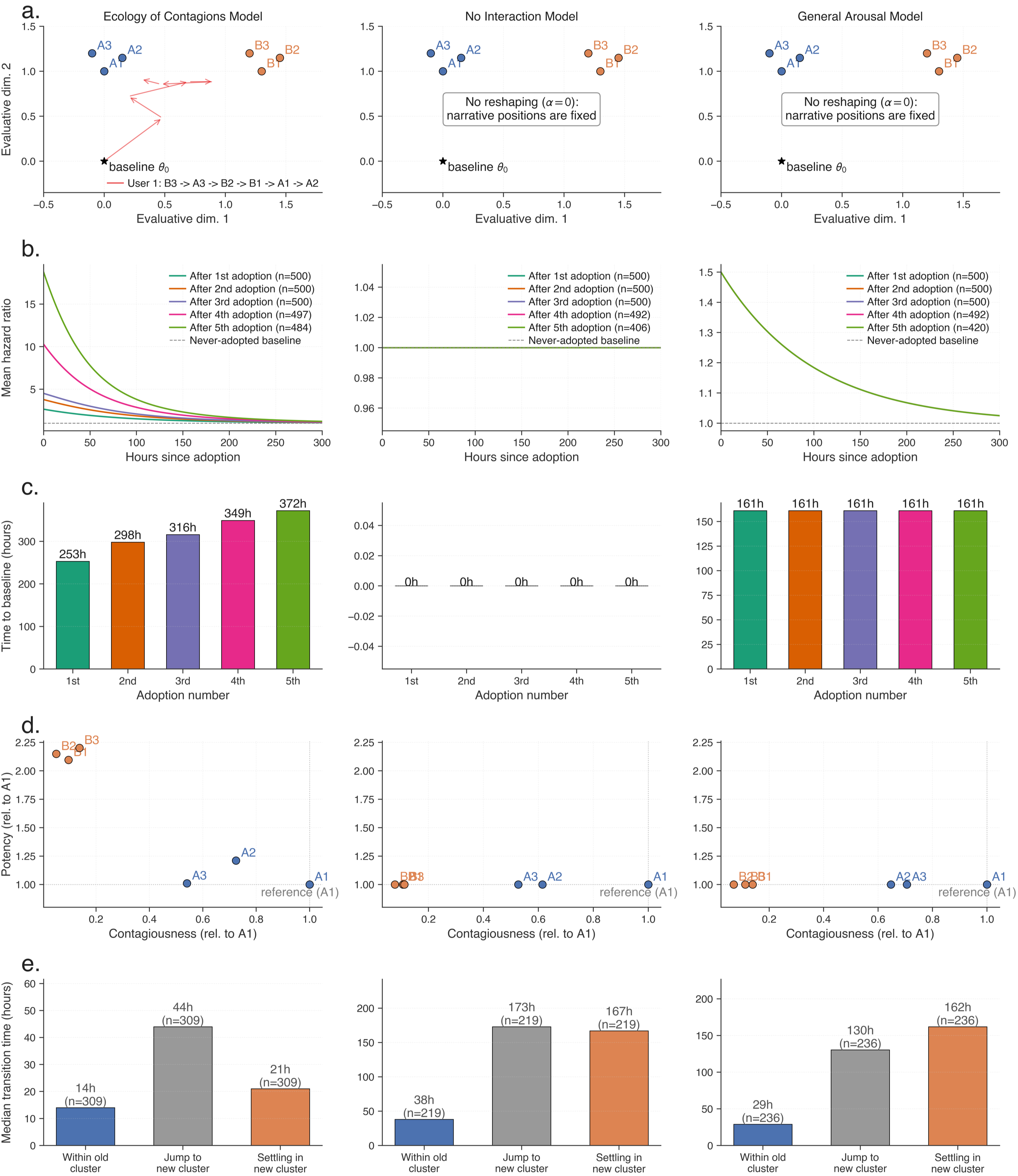}
    \caption{\textbf{Illustrative predictions of the three minimal models.}
    Columns compare the \textit{Ecology of Contagions Model} ($\alpha>0$, $\beta=0$),
    the \textit{No Interaction Model} ($\alpha=0$, $\beta=0$), and
    the \textit{General Arousal Model} ($\alpha=0$, $\beta>0$). The simulation is illustrative rather than fit to data. The \textit{Ecology of Contagions Model} produces structured descent because adoption shifts the user's position in belief system space toward the adopted conspiracy theory, followed by gradual recovery toward baseline. The \textit{No Interaction Model} produces repeated adoption without descent because adoption does not change later susceptibility. The \textit{General Arousal Model} produces descent without adoption specific semantic reshaping because adoption raises susceptibility uniformly across conspiracy theories and this excess susceptibility decays over time.}
    \label{fig:appendix_toy_model}
\end{figure}

Figure~\ref{fig:appendix_toy_model} reports simulations from the three minimal models introduced in the main text: the \textit{No Interaction Model}, the \textit{General Arousal Model}, and the \textit{Ecology of Contagions Model}. These simulations are illustrative rather than fit to the Twitter data. Their purpose is mechanism discrimination: to show which qualitative patterns follow from different assumptions about how prior adoption changes later susceptibility. The three models share the same belief system space, conspiracy theory locations, user population, baseline hazard and simulation horizon. They differ only in the after adoption mechanisms governed by \(\alpha\) and \(\beta\).

Each simulation contains 500 independent users. There is no social network, no peer exposure, no Hawkes repeat posting process and no seeding. All users begin with no adopted conspiracy theories, and all six conspiracy theories are initially in the user's risk set. Users are initialized at the same baseline position in a two dimensional belief system space, \(\theta_i^0=(0,0)\). The six conspiracy theories are placed in two nearby semantic regions. The first contains \(A_1=(0.00,1.00)\), \(A_2=(0.15,1.15)\) and \(A_3=(-0.10,1.20)\). The second contains \(B_1=(1.30,1.00)\), \(B_2=(1.45,1.15)\) and \(B_3=(1.20,1.20)\). Simulations use a one hour time step, a maximum horizon of 1,000 hours and a fixed random seed.

At each hour, the model computes the adoption hazard for every conspiracy theory that the user has not previously adopted:
\[
h_{ic}(t)
=
\mu A_i(t)
\exp\left(
-\frac{\|\sigma_c - \theta_i(t)\|^2}{\rho}
\right),
\]
with \(\mu=0.05\) and \(\rho=0.8\). The total hazard is the sum of these hazards over all conspiracy theories still in the user's risk set. The simulation converts this continuous time hazard into a one hour event probability,
\[
p_i(t)=1-\exp\left(-\Delta t\sum_{c \in R_i(t)} h_{ic}(t)\right),
\]
where \(R_i(t)\) is the set of conspiracy theories not yet adopted by user \(i\) and \(\Delta t=1\) hour. If an adoption occurs, the adopted conspiracy theory is sampled in proportion to its hazard and then removed from that user's future risk set. A user stops being simulated once all six conspiracy theories have been adopted or the horizon of 1,000 hours is reached.

The \textit{No Interaction Model} sets \(\alpha=0\) and \(\beta=0\). Adoption records an event but does not alter the user's belief system state and does not change susceptibility to the remaining conspiracy theories. In this model, \(A_i(t)=1\) for all users and times, and later adoption is governed only by the fixed distances between the baseline state and the conspiracy theories that remain in the risk set. The model can produce repeated adoption and clustered first adoption because users remain at risk after each adoption and some conspiracy theories begin closer to the baseline state, but it cannot produce descent because adoption itself has no effect on later susceptibility.

The \textit{General Arousal Model} sets \(\alpha=0\) and \(\beta>0\). Adoption leaves the user's belief system state unchanged, but it temporarily raises susceptibility to all remaining conspiracy theories uniformly:
\[
A_i(t)
=
1 + \beta \exp[-\gamma(t-t_a)].
\]
In the illustrative simulation, \(\beta=0.5\), \(\gamma=0.01\) per hour and \(t_a\) denotes the user's most recent adoption time. Equivalently, the implementation resets excess susceptibility to 0.5 after each adoption and then decays it each hour by \(\exp(-1/100)\). This arousal term is not accumulated across earlier adoptions. Because this multiplier applies equally to all remaining conspiracy theories, the model creates a temporary period of elevated adoption risk without making semantically nearby conspiracy theories especially likely because of what was just adopted.

The \textit{Ecology of Contagions Model} sets \(\alpha>0\) and \(\beta=0\). Adoption changes the user's belief system state. When user \(i\) adopts conspiracy theory \(c\) at time \(t_a\), the state is immediately reshaped toward the adopted conspiracy theory:
\[
\theta_i(t_a^+)
=
(1-\alpha)\theta_i(t_a^-)
+
\alpha\sigma_c.
\]
In the illustrative simulation, \(\alpha=0.4\). The simulation records \(\theta_i(t_a^+)\) immediately after this reshaping and before the recovery update at the end of the same simulation hour. After reshaping, the state relaxes toward the user's baseline state:
\[
\frac{d\theta_i}{dt}
=
k(\theta_i^0 - \theta_i),
\]
with \(k=0.01\). The simulation implements this recovery equation with the one hour update
\[
\theta_i(t+\Delta t)=\theta_i(t)+k(\theta_i^0-\theta_i(t))\Delta t.
\]
Thus adoption creates a temporary but persistent movement in belief system space. Conspiracy theories near the adopted conspiracy theory become more likely while the belief system state remains displaced, and this effect weakens as recovery pulls the user back toward baseline.

The rows of Figure~\ref{fig:appendix_toy_model} display diagnostics from these simulations. The first row shows the belief system space, the conspiracy theory locations, the common baseline state and example user trajectories. In the \textit{Ecology of Contagions Model}, the plotted trajectory uses the recorded states immediately after each reshaping event. In the \textit{No Interaction Model} and the \textit{General Arousal Model}, semantic position remains fixed because \(\alpha=0\). The second row shows the average hazard ratio for remaining conspiracy theories after prior adoption, relative to the same user's baseline state. For the \textit{Ecology of Contagions Model} and the \textit{No Interaction Model}, these curves are computed by starting from the recorded state after an adoption and then applying recovery deterministically with no further stochastic adoptions. For the \textit{General Arousal Model}, the curve is \(1+\beta\exp[-\gamma(t-t_a)]\). The third row reports the corresponding interaction window, defined as the first time at which the mean hazard ratio falls below 1.1 times baseline. The fourth row shows the contagiousness and potency tradeoff. Contagiousness is measured as first adoption frequency relative to \(A_1\). Potency is computed only from first adoption events and is defined as the mean later hazard increase induced by adopting that conspiracy theory, again relative to \(A_1\). The fifth row summarizes the settler pattern using sequences of the form old, old, new, new, where the new semantic region has not been visited earlier by that user. It compares the median waiting time before the transition, during the transition and after settling in the new region.

Together, the panels show that the \textit{Ecology of Contagions Model} uniquely produces belief system movement, persistent interaction windows, variation in potency across conspiracy theories and the settler pattern under this shared initialization.

\paragraph{Interpretation of the model comparison.}

These simulations are illustrative rather than fitted tests of private belief change. All three models use the same belief system space, so the semantic positions of users and conspiracy theories can generate some semantic pattern even when adoption does not reshape the user. Under the shared initialization used here, the Ecology of Contagions Model is the only model among the alternatives considered that reproduces the full joint pattern of elevated later adoption, semantic concentration and the settler effect. This result shows that belief system reshaping provides a minimal mechanism capable of generating the observed combination of patterns. It does not establish that this is the only possible mechanism or directly observe changes in private belief.

\section{Simulation validation}\label{apx:sec_simulation_validation}

The following figures document the fidelity with which the agent-based simulation reproduces the empirical spreading dynamics. Adoption probabilities inherit the fitted sequential event-history models, and repeat posting follows the fitted Hawkes self-exciting process. No diffusion parameters are hand-tuned.

\begin{figure}[H]
    \centering
    \includegraphics[width=\textwidth]{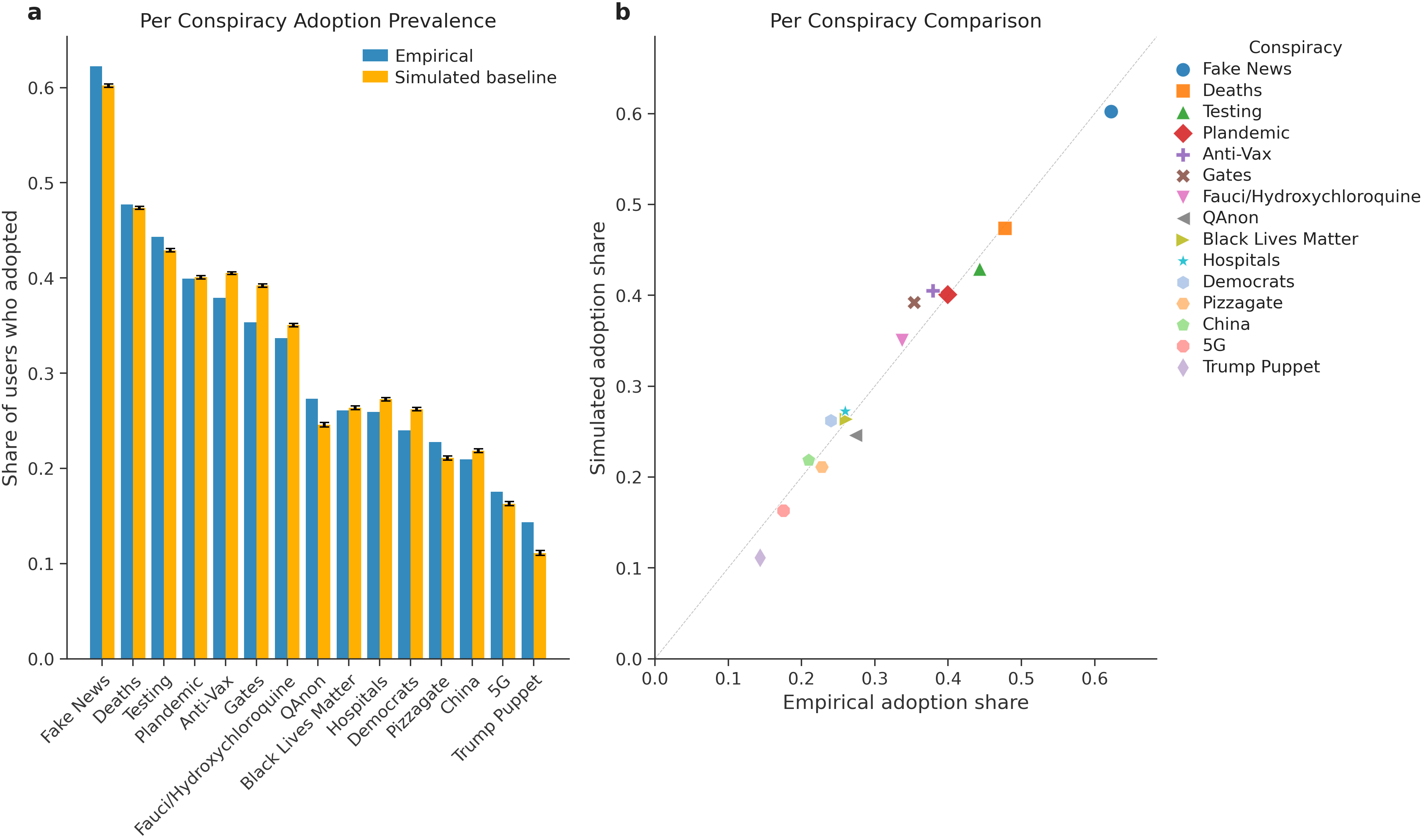}
    \caption{\textbf{Per-contagion adoption prevalence: empirical versus simulated.}
    \textit{Left:} Share of users who adopted each analytic contagion in the empirical data and in the baseline simulation, with simulated means and 95\% CI across runs. \textit{Right:} Scatter plot of empirical versus simulated adoption shares. The simulation reproduces the rank ordering and magnitudes of per-contagion adoption rates without post-hoc tuning.}
    \label{fig:appendix_prevalence}
\end{figure}

\begin{figure}[H]
    \centering
    \includegraphics[width=0.6\textwidth]{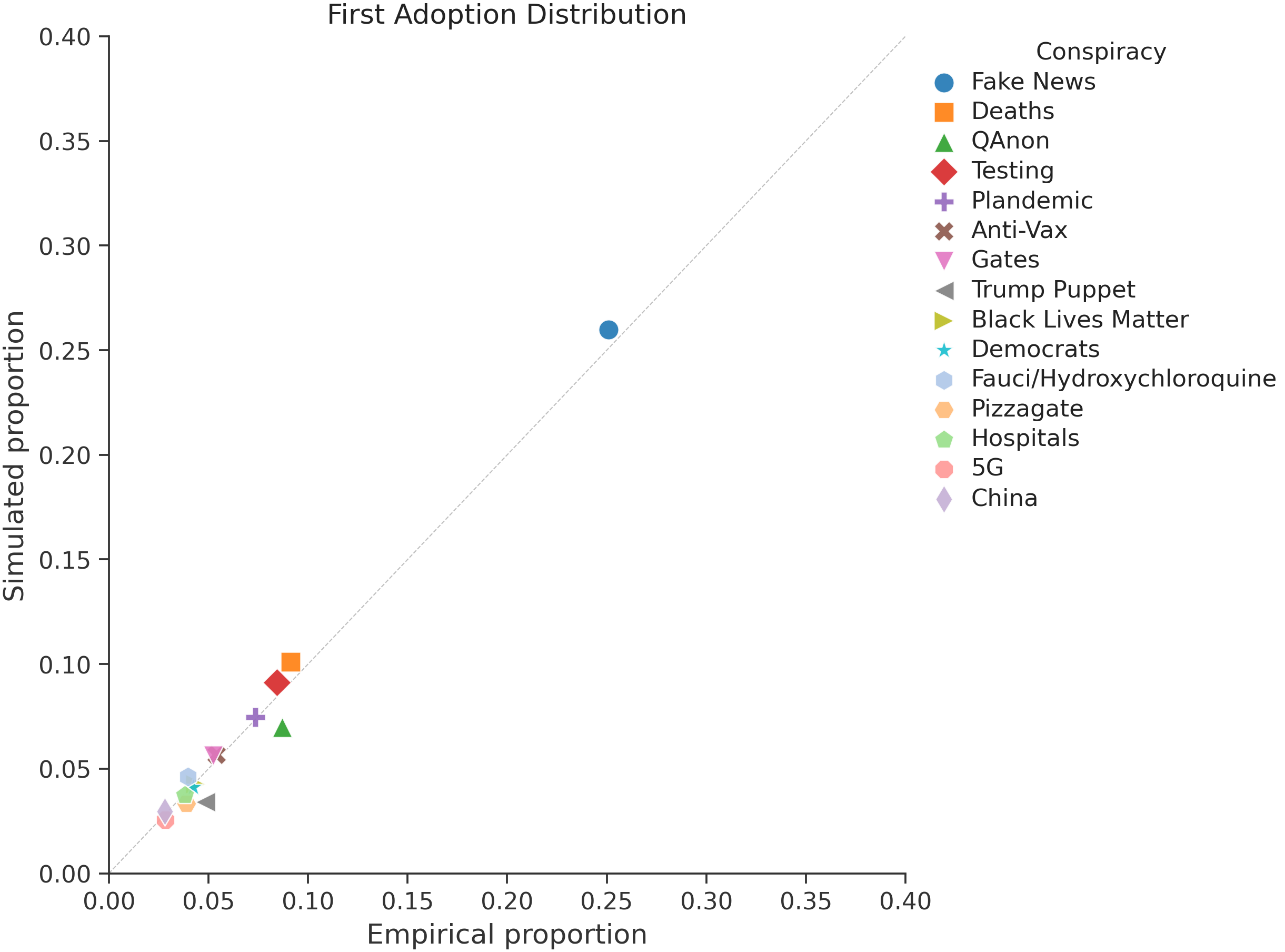}
    \caption{\textbf{First-adoption entry-point distribution: empirical versus simulated.}
    Each point represents one conspiracy theory. The x-axis shows the empirical proportion of users who adopted that contagion first; the y-axis shows the simulated proportion in the baseline simulation. The simulation reproduces which contagions function as common entry points into the adoption sequence.}
    \label{fig:appendix_first_adoption}
\end{figure}

\begin{figure}[H]
    \centering
    \includegraphics[width=\textwidth]{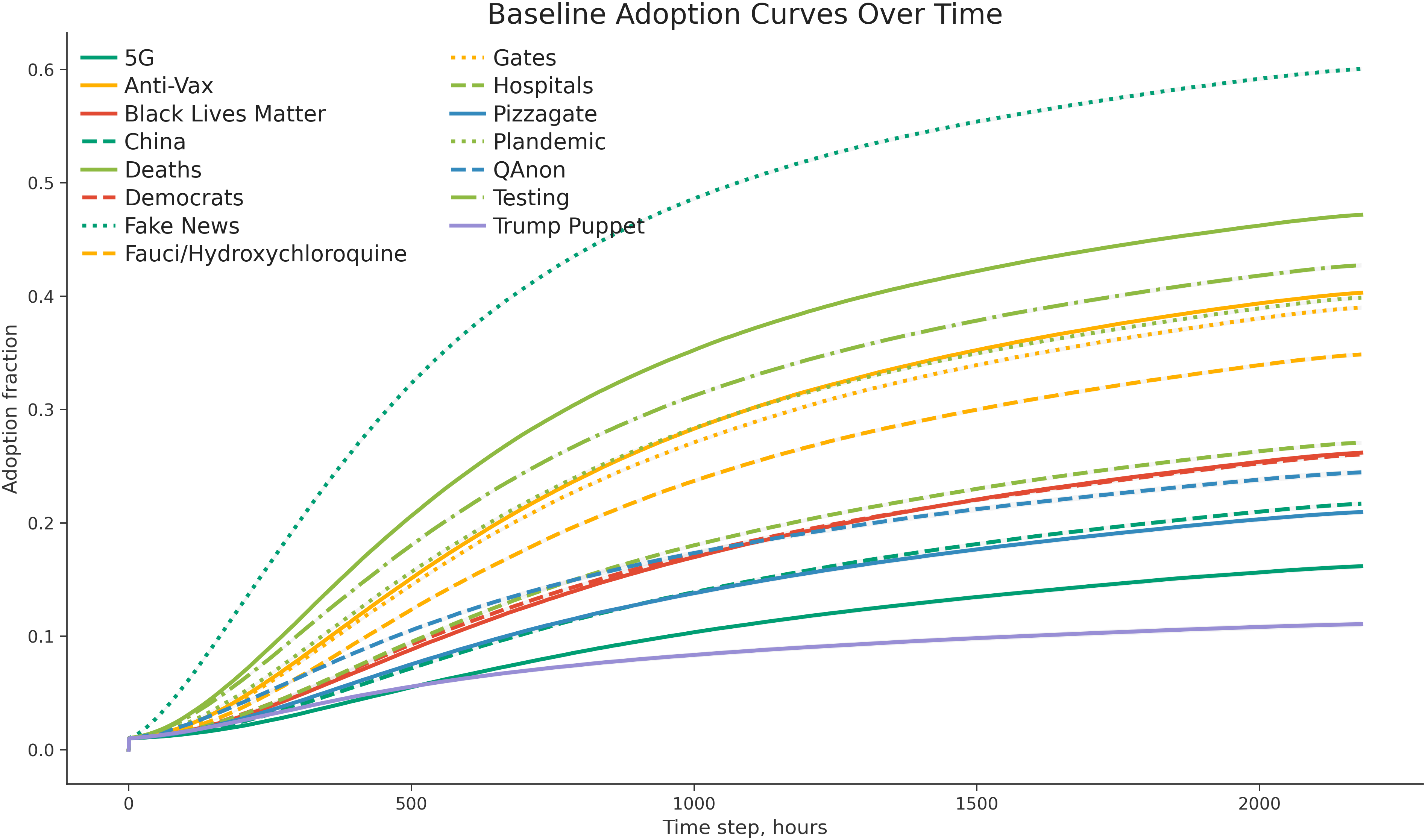}
    \caption{\textbf{Baseline diffusion curves.}
    Adoption fraction over time for each analytic contagion in the baseline simulation, shown as means across runs. The curves show the simulated temporal development of adoption under the fitted interacting-contagion dynamics.}
    \label{fig:appendix_diffusion}
\end{figure}

\begin{figure}[H]
    \centering
    \includegraphics[width=\textwidth]{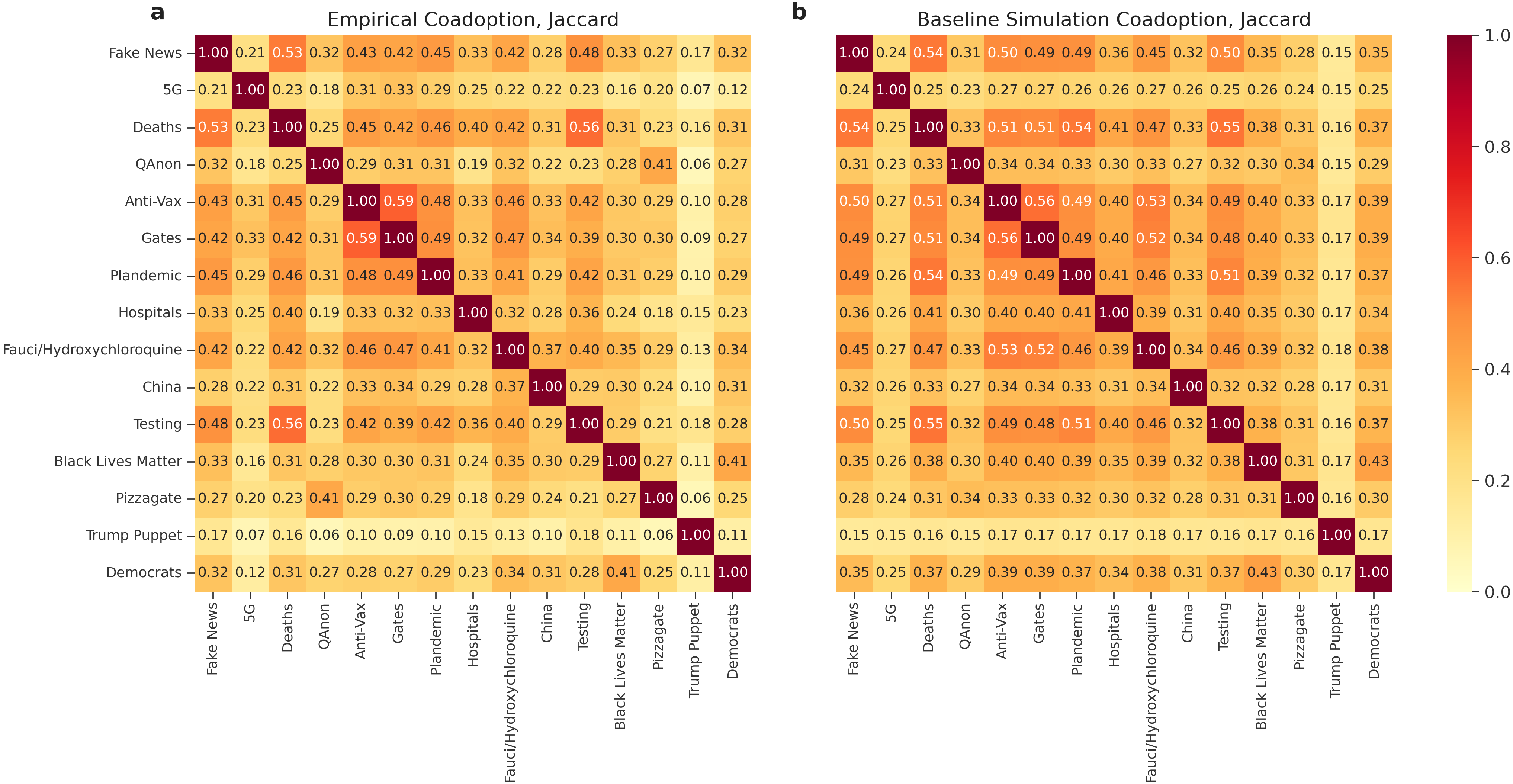}
    \caption{\textbf{Co adoption structure: empirical and simulated Jaccard matrices.}
    \textit{(a)} Pairwise Jaccard similarity of analytic contagion co adoption across users in the empirical data. \textit{(b)} The same matrix computed from the baseline simulation, averaged across runs. The simulation reproduces which contagions tend to be adopted together without parameter tuning beyond the fitted event history and posting models.}
    \label{fig:appendix_coadoption_matrices}
\end{figure}

\begin{figure}[H]
    \centering
    \includegraphics[width=0.72\textwidth]{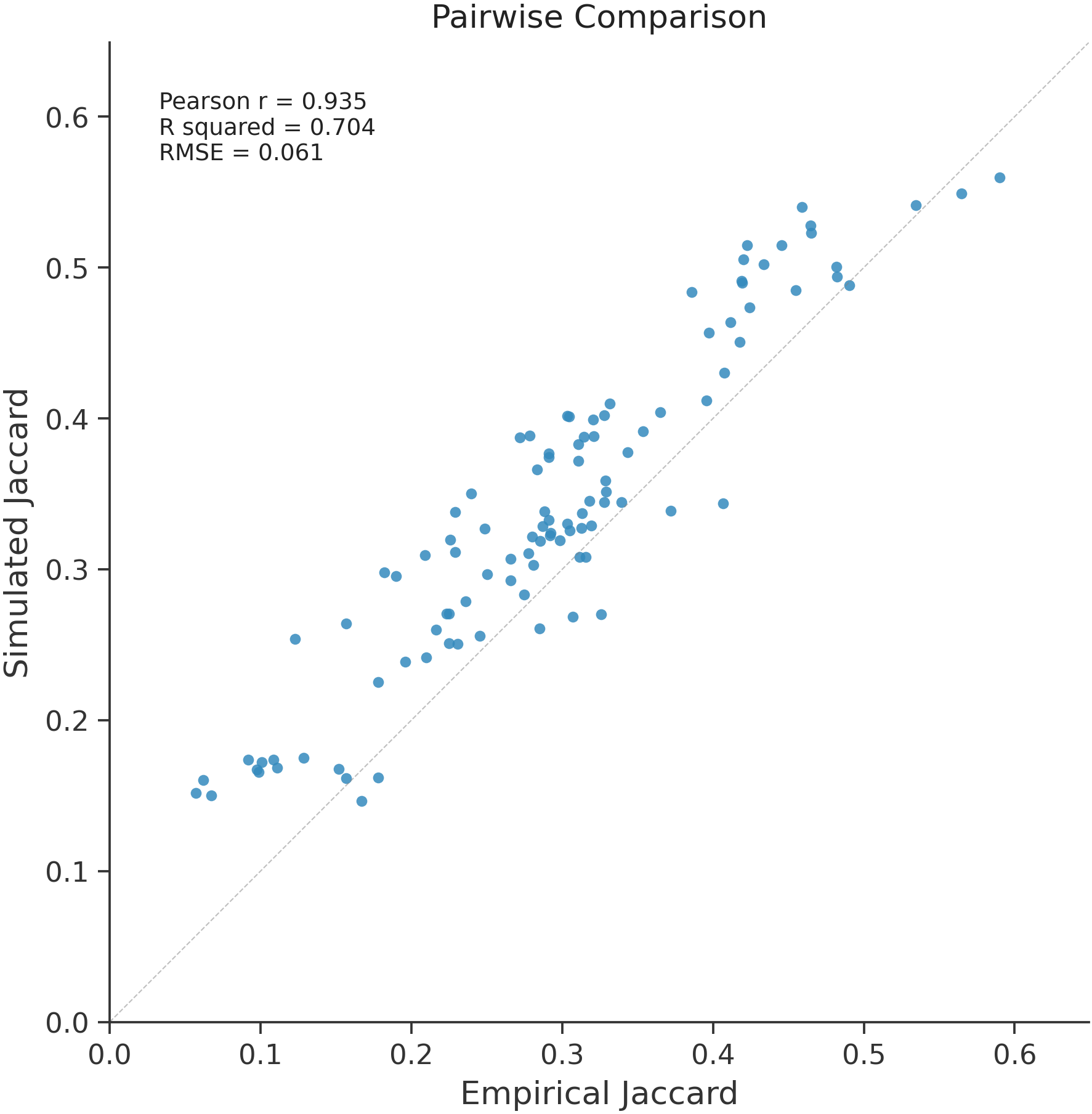}
    \caption{\textbf{Empirical and simulated co adoption structure.} Each point represents one of the 105 unique pairs among 15 analytic contagions. Jaccard similarity is the number of users adopting both contagions divided by the number adopting either. The x axis shows empirical similarity among 4,516 human users, and the y axis shows mean similarity across 100 baseline simulation runs on the same network. The grey diagonal marks exact agreement. Pearson \(r=0.935\) measures linear association. Agreement \(R^2=0.704\) is defined as \(1-\sum_p(J^{\mathrm{emp}}_p-J^{\mathrm{sim}}_p)^2/\sum_p(J^{\mathrm{emp}}_p-\bar J^{\mathrm{emp}})^2\), over contagion pairs \(p\). RMSE is \(0.061\). Simulated similarities are higher on average than empirical similarities, partially explaining why agreement \(R^2\) is lower than Pearson \(r^2=0.874\).}
    \label{fig:appendix_coadoption_scatter}
\end{figure}

\section{Policy-counterfactual diagnostics}

\begin{figure}[H]
    \centering
    \includegraphics[width=\textwidth]{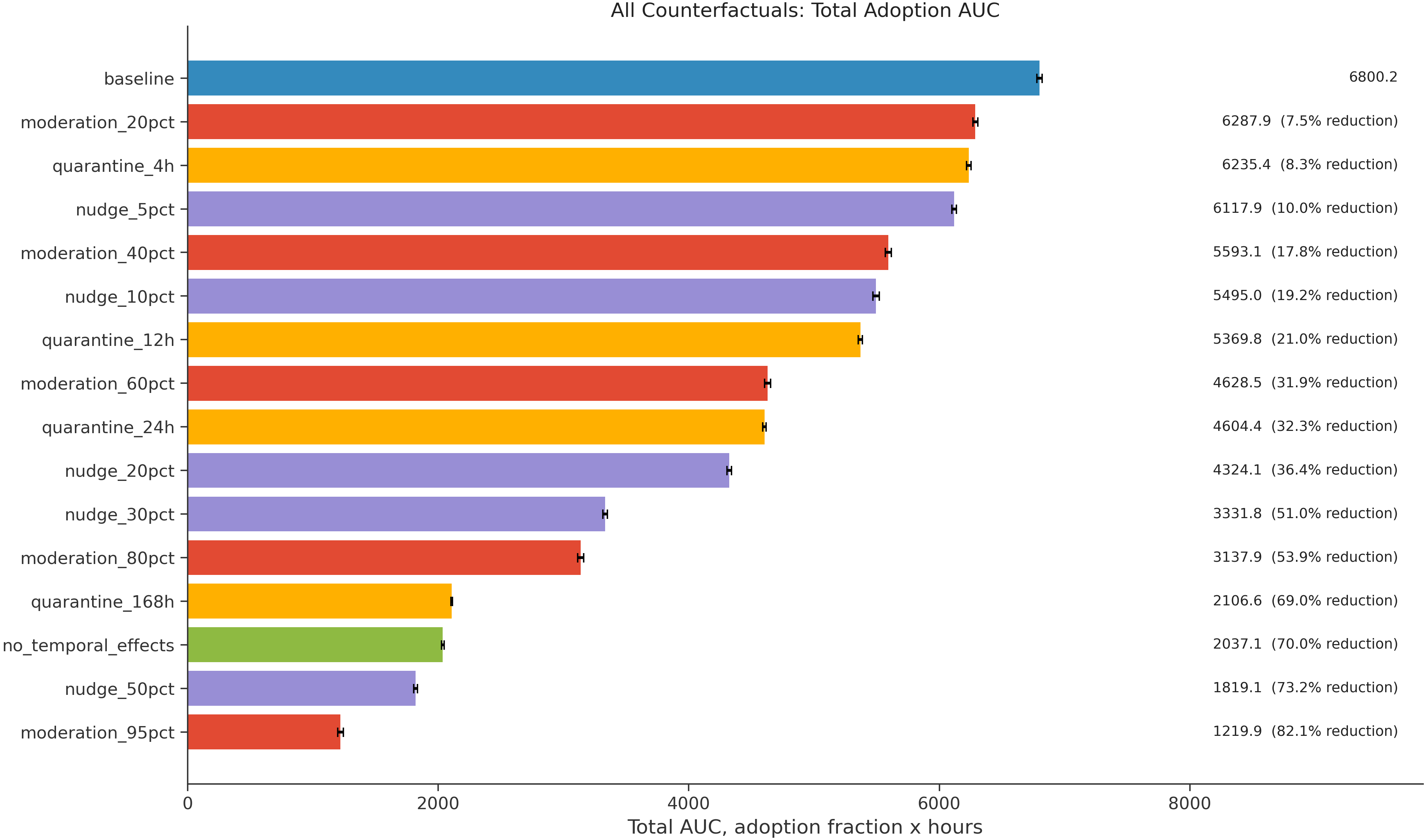}
    \caption{\textbf{All counterfactual scenarios ranked by total area under the adoption curve.} Horizontal bars show the total area under the adoption curve for each scenario, with error bars showing 95\% bootstrap CIs. Percentage reductions relative to the baseline are annotated. The ranking shows that reputation nudges achieve comparable reductions to shadow banning (called moderation here) at substantially lower intervention intensities because they prevent the adoption event itself rather than only suppressing visibility after adoption.}
    \label{fig:appendix_auc_ranking}
\end{figure}

\section{Repeat sharing validation}\label{sec:repeated_sharing_appendix}

The sequential hazard models determine when a user adopts a conspiracy theory for the first time. The repeat sharing process determines what happens after that adoption. In the simulation, adoption creates the first public share for that user and conspiracy theory. Subsequent shares are generated by a pooled Hawkes process, which allows sharing to be bursty: each share temporarily increases the rate of additional shares about the same conspiracy theory by the same user.

For each user and conspiracy theory, we extracted the empirical sharing sequence after the first observed share. Times were shifted so that the first share occurs at \(t=0\). The Hawkes process is fitted to repeat sharing sequences pooled across conspiracy theories, with shared parameters for all user and theory histories. The likelihood is conditioned on the first share, so the first share is treated as the adoption event and is not itself generated by the repeat sharing process. All 21,683 user and conspiracy theory histories contribute to parameter fitting; 12,850 histories contain at least one repeat share, yielding 96,325 repeat share events. When estimating the likelihood, the observation window extends from the first share to the end of the study period, so silence after the last observed share contributes to the compensator.

For user \(i\) and conspiracy theory \(c\), the fitted Hawkes intensity is
\[
\lambda_{ic}(t)
=
\mu
+
\alpha\beta
\sum_{t_j<t}
\exp[-\beta(t-t_j)],
\]
where \(t_j\) are prior shares by the same user about the same conspiracy theory. Here \(\mu\) is the baseline repeat sharing rate, \(\alpha\) is the excitation parameter and \(\beta\) is the decay rate. The fitted pooled parameters used in the simulation are \(\mu=0.0004435\), \(\alpha=0.7176\) and \(\beta=0.009533\). The corresponding mean excitation duration is \(1/\beta=104.9\) hours. Because \(\alpha<1\), the fitted process is subcritical.

In the agent based simulation, the Hawkes process is evaluated at each one hour step after adoption. The adoption share is already recorded at the adoption time, so repeat sharing is skipped in that same hour. For later hours, the implementation integrates the Hawkes intensity over the next hour,
\[
m_{ic}(t)
=
\mu\Delta t
+
\alpha(1-\exp[-\beta\Delta t])
\sum_{t_j<t}
\exp[-\beta(t-t_j)],
\]
with \(\Delta t=1\) hour, and draws the number of repeat shares from a Poisson distribution with mean \(m_{ic}(t)\). Events older than \(21/\beta\) are ignored when evaluating the sum, retaining effectively all of the exponential kernel mass. These repeat shares update the user's sharing history and, when visible, enter neighbours' 14 day exposure histories. Thus the Hawkes process does not determine adoption directly. It determines the temporal clustering of shares after adoption, which in turn affects later peer exposure.

\begin{figure}[H]
    \centering
    \includegraphics[width=\textwidth]{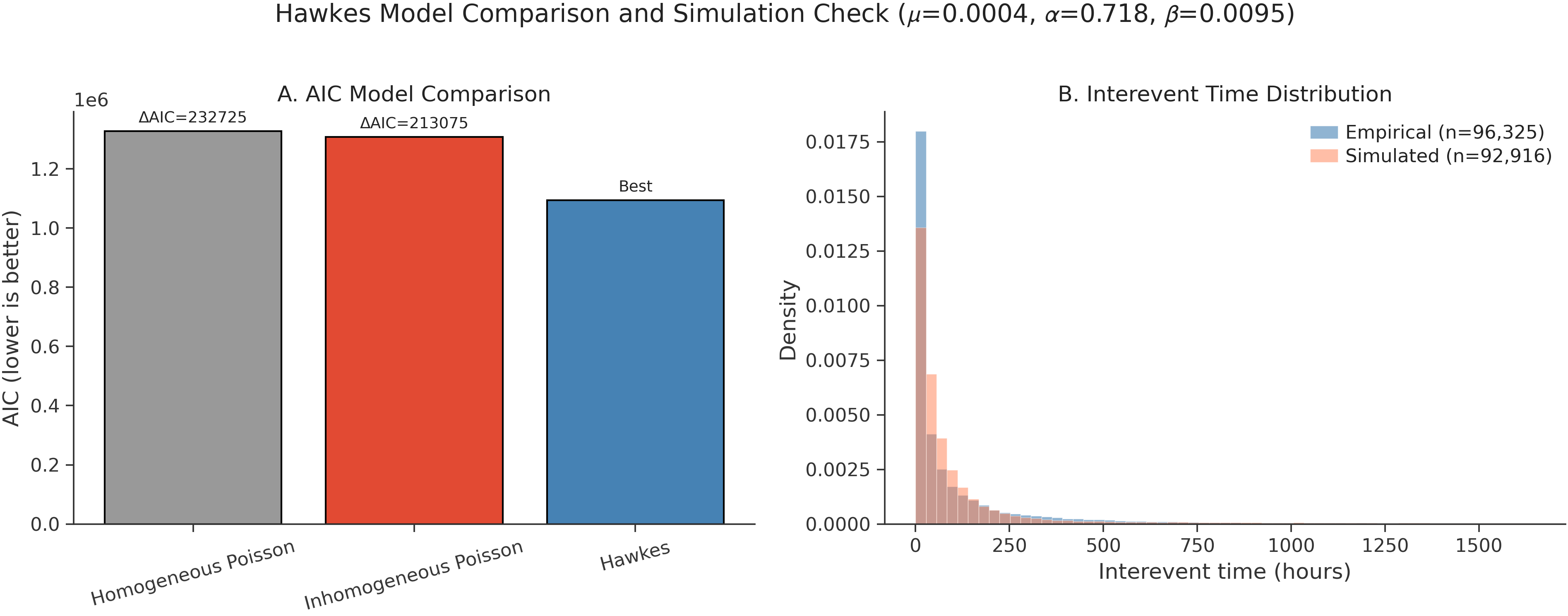}
    \caption{\textbf{Hawkes process goodness of fit diagnostics.}
    Two panel validation of the pooled Hawkes repeat sharing process fitted to empirical sharing sequences. \textit{A}: AIC model comparison against homogeneous and inhomogeneous Poisson alternatives. \textit{B}: Empirical and Hawkes simulated distributions of times between repeat shares, with each simulated history using the observation window of its empirical counterpart and conditioned on an initial event at \(t=0\). The diagnostics support using a self exciting Hawkes process, rather than either Poisson alternative, to capture bursty repeat sharing after adoption.}
    \label{fig:appendix_hawkes}
\end{figure}

We validate this repeat sharing process in Figure~\ref{fig:appendix_hawkes}. The current two panel diagnostic combines an information criterion model comparison with a simulation check. Panel A compares the fitted Hawkes process with homogeneous and inhomogeneous Poisson alternatives using AIC. Panel B compares the empirical distribution of times between repeat shares with the distribution generated by simulations from the fitted Hawkes process.

Panel A compares the Hawkes model with two Poisson alternatives using AIC. The homogeneous Poisson model has a single constant rate. The inhomogeneous Poisson model has intensity
\[
\lambda(t)=a\exp[-bt]+c,
\]
which allows repeat sharing to decline with time since first share but does not allow realized shares to excite later shares. The AIC comparison shows that the fitted Hawkes process is strongly preferred over the homogeneous and inhomogeneous Poisson alternatives for repeated sharing sequences. This supports using a self exciting process to capture bursty repeat sharing after adoption. The fit includes all 21,683 user and conspiracy theory histories, including histories with no repeat share, whose subsequent silence contributes through the compensator. This is important for the simulation because repeat shares are what enter neighbours' 14 day exposure histories and can therefore generate additional opportunities for adoption.

Panel B compares the pooled empirical distribution of interevent times following the initial adoption share with the corresponding distribution from continuous time simulations of the fitted Hawkes model. Each simulated history uses the exact observation window of one empirical history and is conditioned on an initial adoption event at \(t=0\).

\section{Exposure dose-response robustness}

Figure~\ref{fig:hazard_panel}a in the main text shows a simplified two-line exposure dose-response: first adoption and second-or-later adoption. The figure below reports the full per-model breakdown across adoption stages.

\begin{figure}[H]
    \centering
    \includegraphics[width=\textwidth]{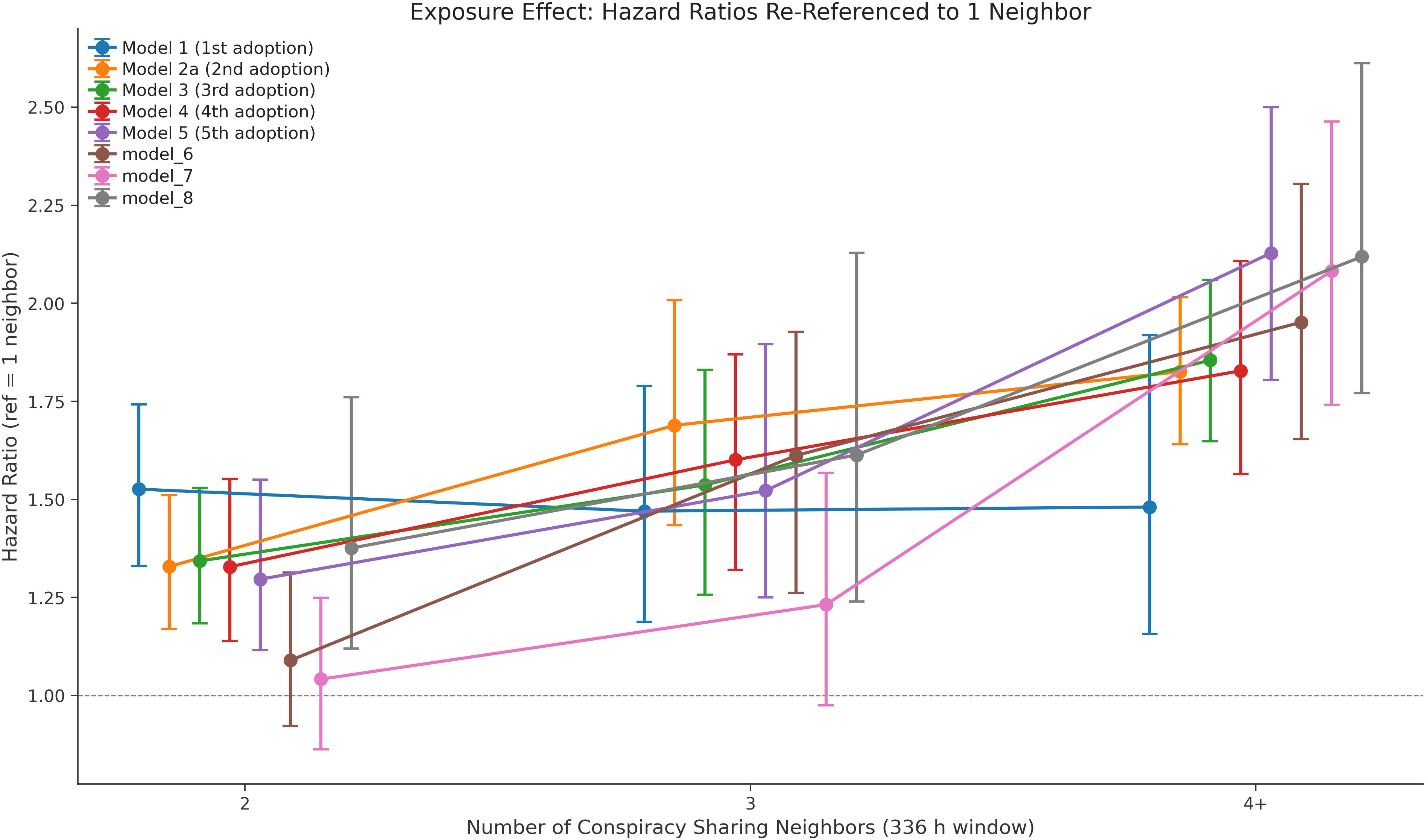}
    \caption{\textbf{Full social-reinforcement dose-response across adoption stages.}
    Hazard ratios for conspiracy-theory adoption as a function of the number of sharing neighbors, shown separately across adoption-stage models. The first-adoption curve is distinct from later-adoption curves. Subsequent adoption stages show stronger response to multiple exposed neighbors than first adoption, while variation among later stages is comparatively smaller.}
    \label{fig:appendix_full_doseresp}
\end{figure}

\section{Interpretation of the sequential hazard baselines}

Figure~\ref{fig:hazard_panel} compares baseline hazards from the sequential Cox models described in the Methods. Model 1 estimates entry into the adoption sequence, while each later model estimates the next adoption among users who have reached the preceding adoption depth. The users, available conspiracy theories and covariate distributions therefore change across models as users progress through the observed adoption sequence.

The models were estimated using \texttt{lifelines}, which centers covariates at their means within the data used to fit each model. For adoption depth \(k\), the fitted hazard is

\[
h_k(t \mid \mathbf{x})
=
h_{0k}(t)
\exp\left[
(\mathbf{x}-\bar{\mathbf{x}}_k)^{\top}
\boldsymbol{\beta}_k
\right],
\]

where \(\bar{\mathbf{x}}_k\) is the mean covariate profile for the observations included in model \(k\). The baseline hazard \(h_{0k}(t)\) therefore describes the fitted hazard at the mean covariate profile of the users and conspiracy theories at risk at that adoption depth.

We use these stage specific centered baselines because our theoretical interest is in how the adoption process unfolds as users move through successive stages. Each stage represents a different transition. Users have accumulated different adoption histories, previously adopted conspiracy theories have left the risk set, peer exposure has continued to change and the population reaching the next stage has been selected through the preceding adoption process. These changes are not treated as incidental differences to be removed. They are part of the ecology through which later adoption occurs.

Evaluating every model at a common covariate profile would instead describe a hypothetical population whose measured characteristics remain fixed across the adoption sequence. That comparison is useful for isolating the role of the measured covariates, but it does not represent the populations observed at each adoption depth. It may also evaluate later models at combinations of covariates that are uncommon at those stages. We therefore use the model specific centered baselines as our primary descriptive comparison.

The resulting estimates describe how the fitted adoption process changes across the populations that actually reach successive adoption depths. They retain changes in observed composition that emerge through prior adoption, exposure and censoring. They are not intended as controlled comparisons of the same fixed individual at different depths. Common profile comparisons preserve the broad elevation in later adoption hazard, although the precise depth specific estimates vary with the chosen reference profile.

The recovery windows in Figure~\ref{fig:hazard_panel} are defined as the time at which each later adoption baseline returns to the first adoption baseline. They provide a common descriptive measure of the duration of elevated later adoption hazard across the fitted models. Because the models refer to different adoption stages and centered covariate profiles, these windows describe intersections between the fitted stage specific baselines. They should not solely be interpreted as direct measurements of when an individual private belief system returns to its state before adoption.

Uncertainty in the baseline comparisons and recovery windows was estimated using the user level bootstrap described in the Methods.
\end{document}